\documentclass[conference]{IEEEtran}

\IEEEoverridecommandlockouts

\usepackage{cite}
\usepackage{amsmath,amssymb,amsfonts}
\usepackage{algorithm}
\usepackage{algpseudocode}
\usepackage{graphicx}
\usepackage{textcomp}
\usepackage{xcolor}
\usepackage{hyperref}
\usepackage{physics}
\usepackage{subcaption}
\usepackage[dvipsnames,table]{xcolor}
\usepackage{booktabs}
\usepackage{multirow}
\usepackage{enumitem}
\setlist[enumerate]{itemsep=5pt, topsep=5pt}
\setlist[itemize]{itemsep=5pt, topsep=5pt}
\usepackage[table]{xcolor}

\definecolor{summarygray}{gray}{0.92}
\usepackage{fancyhdr}
\usepackage{soul}
\def\BibTeX{{\rm B\kern-.05em{\sc i\kern-.025em b}\kern-.08em
T\kern-.1667em\lower.7ex\hbox{E}\kern-.125emX}}

\makeatletter
\renewcommand{\IEEEauthorrefmark}[1]{\textsuperscript{#1}}
\makeatother

\begin{document}

\bstctlcite{BSTcontrol}

\title{Parallel Circuit Execution for Scalable Quantum Computation}

\author{
    \IEEEauthorblockN{
    Avimita Chatterjee,\IEEEauthorrefmark{1}
    W. Michael Brown,\IEEEauthorrefmark{2}
    Siyuan Niu,\IEEEauthorrefmark{3}
    Wibe Albert de Jong,\IEEEauthorrefmark{1}
    Thomas Lubinski,\IEEEauthorrefmark{4,5,6}
    }
\\
\IEEEauthorblockA{\IEEEauthorrefmark{1}\textit{Lawrence Berkeley National Laboratory, Berkeley, CA 94720, USA}}
\IEEEauthorblockA{\IEEEauthorrefmark{2}\textit{NVIDIA, Santa Clara, CA 95051, USA}}
\IEEEauthorblockA{\IEEEauthorrefmark{3}\textit{University of Central Florida, Orlando, FL 32816, USA}}
\IEEEauthorblockA{\IEEEauthorrefmark{4}\textit{Quantum Computing Data, San Rafael, CA 94901, USA}}
\IEEEauthorblockA{\IEEEauthorrefmark{5}\textit{Cascade Quantum, San Rafael, CA 94901, USA}}
\IEEEauthorblockA{\IEEEauthorrefmark{6}\textit{QED-C Technical Advisory Committee -- Standards, Arlington, VA 22209, USA}}
}

\maketitle

\begin{abstract}
Today's quantum processors have tens to hundreds of physical qubits, but reliable execution of arbitrary circuits remains limited to fewer than 30 entangled qubits across hardware modalities.
Building upon prior work, we introduce an error- and topology-aware method for mapping multiple independent circuits onto disjoint regions of a single large-scale QPU for parallel circuit execution.
For applications with many similarly sized circuits, such as observable estimation for Hamiltonian simulation, this approach can reduce billed QPU execution time, with ideal speedup proportional to the number of usable partitions.
We demonstrate the approach on IBM's 156-qubit \texttt{ibm\_boston} processor using standard QED-C benchmark and Hamiltonian based observable-estimation workloads. Compared with standard sequential execution, parallel execution reduces billed execution time by 3.5–5.5$\times$ while retaining 83–92\% of the sequential fidelity.
We further evaluate the scaling of parallel circuit execution using GPU-accelerated classical simulation, distributing measurement circuits across GPUs via MPI. Using CUDA-Q on the NERSC Perlmutter system, we achieve up to 13.8$\times$ speedup on 16 GPUs (86\% parallel efficiency) for an H2 electronic-structure simulation, with scaling evaluated across multiple Hamiltonians and circuit counts.
These results provide an indication of the performance ceiling that parallel execution on future quantum hardware may eventually approach. Both execution modes are implemented as a runtime option within the QED-C Application-Oriented Benchmark suite. Together, the results show that circuit-level parallelism can reduce execution cost on current quantum hardware and simulation time on GPU clusters, with the potential for greater benefits as device quality and qubit counts increase.

\end{abstract}
\vspace{1em}



\pagestyle{fancy}
\renewcommand{\headrulewidth}{0.0pt}
\lhead{}
\rhead{\thepage}
\renewcommand{\footrulewidth}{0.4pt}
\cfoot{}
\lfoot{Parallel Circuit Execution for Scalable Quantum Computation}
\rfoot{\today}

\section{Introduction}
\label{sec:introduction}

\begin{figure*}[t]
    \centering
    \includegraphics[width=0.9\linewidth]{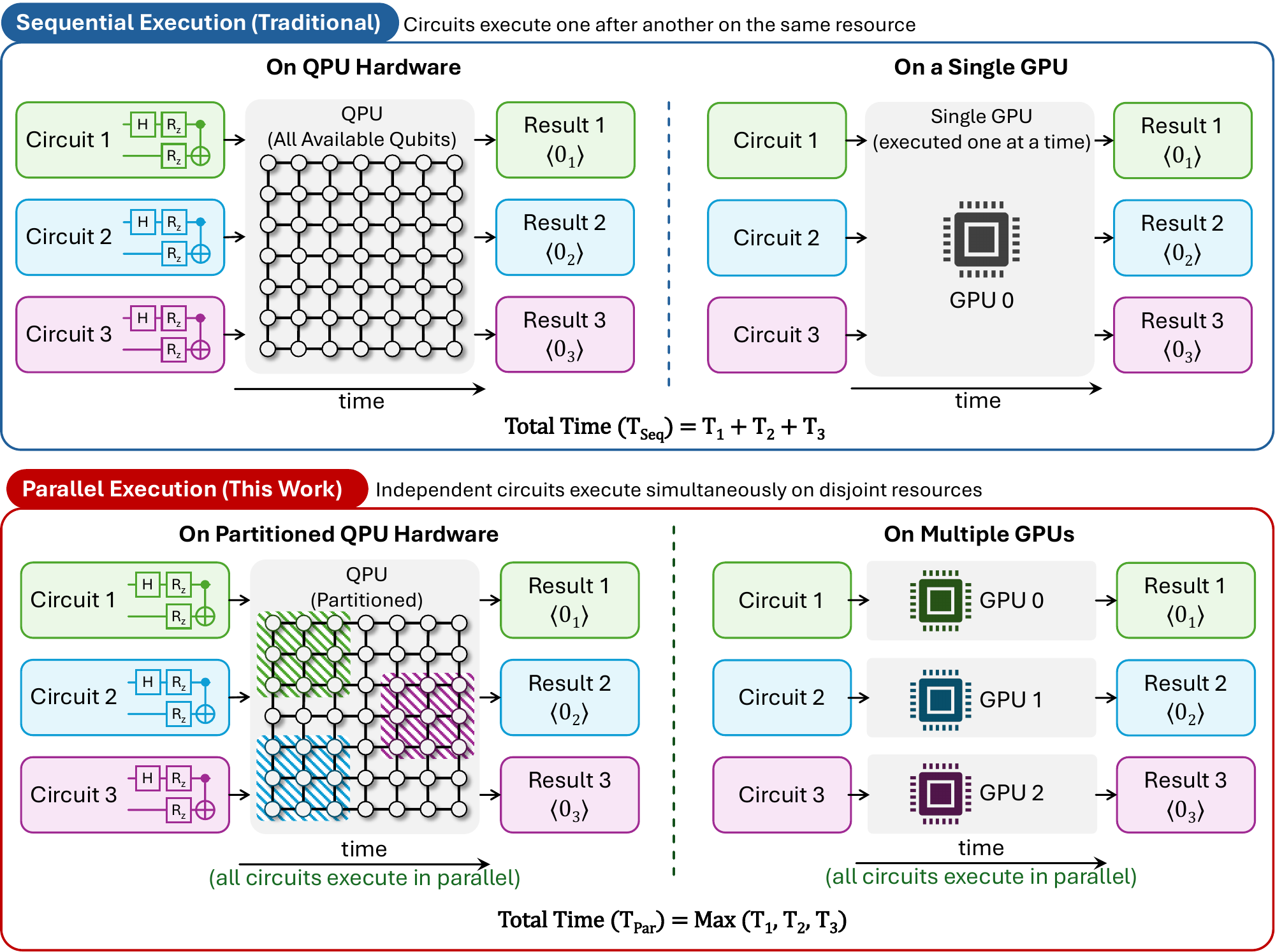}
   \caption{\textbf{Conceptual overview of sequential and parallel circuit execution.}
    Top: Traditional execution, where independent quantum circuits are executed sequentially on a single QPU or GPU, resulting in total execution time equal to the sum of the individual circuit times. Bottom: The parallel execution strategy, where independent circuits are mapped onto disjoint qubit partitions within a single QPU or distributed across multiple GPUs, allowing simultaneous execution. The overall execution time is therefore determined by the longest-running circuit rather than the sum of all circuit times, improving hardware utilization and reducing execution time for workloads consisting of many independent circuits.}
\label{fig:seq_vs_par}
\end{figure*}

Quantum computing has made remarkable progress in a short time. In barely a decade, the field has moved from 5-qubit laboratory demonstrations to processors with over 1,000 physical qubits, accompanied by steady improvements in gate fidelities, coherence times, and software tooling~\cite{doi:10.1073/pnas.1618020114,Garc_a_Mart_n_2018,AbuGhanem_2025,dimolfetta2026quantumcomputingnew,ibmq2023_condor}. Superconducting, trapped-ion, and photonic platforms have each demonstrated capabilities that were considered out of reach only a few years ago, and the first claims of computational advantage over classical systems have generated significant interest across the scientific community~\cite{Ransford2026A9T,Arute2019QuantumSU,Wu_2021,Zhong_2020}.

However, there is a gap between the number of qubits available on a device and the number that can be used reliably in a single computation. Current superconducting processors offer 50 to 150 physical qubits~\cite{osti_2973298,Morvan_2024}, but noise limits reliable circuit execution to roughly 10 to 15 entangled qubits~\cite{Mooney_2021,Cao2023GenerationOG}; trapped-ion systems reach perhaps 30~\cite{Moses_2023,Chen_2024}. Fault-tolerant logical qubits may close this gap through error correction, but remain years away and will themselves be limited in number when they first arrive~\cite{Preskill_2018,huang2025vastworldquantumadvantage,NAP25196,qec_threshold2024,suppress_errors_2023}. Meanwhile, the majority of a processor's qubits sit idle during any given circuit execution~\cite{NicoKatz2024CanQC,DeCross2022QubitReuseCW}. This is likely to persist through several hardware generations and represents a substantial inefficiency in how quantum resources are used today.

Here, we introduce a method that can utilize this idle capacity. Building upon prior work in multiprogramming quantum computers (MPQC)~\cite{qiskit_experiments,das2019case, liu2021qucloud, niu2023enabling,Romo2026MultiQMN,Cao2025AnOR}, we implement a topology-aware partition mapping algorithm that efficiently partitions a quantum processor's qubits into low-error execution zones. 
By running multiple independent circuits simultaneously, billed QPU execution time can be reduced, with the ideal reduction determined by the number of circuits and usable partitions.
Fig.~\ref{fig:seq_vs_par} illustrates the central concept of this work. Independent quantum circuits that would traditionally execute sequentially on a single computational resource are instead mapped onto disjoint QPU partitions or multiple GPUs, allowing simultaneous execution and reducing overall execution time.

The approach targets workloads containing many independent circuits of similar size. We evaluate both standard QED-C benchmarks and observable-estimation workloads, where Hamiltonian Pauli terms are grouped into independently executable measurement circuits. The number of such groups can range from 2 for a simple lattice model to over 1,000 for a molecular electronic structure Hamiltonian, making this a natural setting for circuit-level parallelism. 

To develop and validate the parallel execution strategy, we first use GPU-accelerated classical simulation as a controlled reference environment~\cite{cudaq,brown2025multigpuquantumcircuitsimulation}. On the NERSC Perlmutter system, measurement circuits are distributed across up to 16 GPUs via CUDA-Q~\cite{cudaq} and MPI~\cite{Perlmutter,MPI}. There, we achieve speedups that scale with the number of independent circuit groups: $1.9\times$ for a 2-group Hamiltonian (TFIM), $8.2\times$ for a 9-group Hamiltonian (Bose-Hubbard), and $13.8\times$ for a 1,251-group Hamiltonian (H2 electronic structure), the last representing 86\% of the theoretical maximum. GPU simulation provides 100\% fidelity results at these qubit widths, establishing a performance ceiling against which hardware implementations can be measured.

We then evaluate the strategy on IBM’s 156-qubit \texttt{ibm\_boston} processor~\cite{ibmq2026} using both standard QED-C benchmarks and observable-estimation workloads. Across these experiments, we measure execution-time reduction and the fidelity tradeoffs introduced by partitioned execution.
As an example, for Bose-Hubbard observable estimation at 8 qubits, this yields a $2.8\times$ reduction in billed quantum execution time while retaining 89.5\% of sequential fidelity. In some cases, the parallel result is closer to the exact value, consistent with the error-aware partition mapper selecting favorable regions of the processor.

Parallel execution on current hardware involves tradeoffs between the number and quality of available partitions, circuit fidelity, and the additional setup required for partitioned execution. Some of this setup can be performed in advance and reused across executions, particularly for workloads that repeat with different run-time parameters. As devices grow and improve, more circuits can be executed concurrently, allowing hardware implementations to move closer to the parallel scaling demonstrated here on GPUs.

The GPU and hardware execution modes described in this work are implemented within the QED-C Application-Oriented Benchmark suite~\cite{Lubinski_2023,QEDC-App-Benchmarks,lubinski2024quantumalgorithmexplorationusing,chatterjee2024comprehensivecrossmodelframeworkbenchmarking,niu2025practicalframeworkassessingperformance}. Building on our prior work in Pauli term grouping~\cite{niu2025practicalframeworkassessingperformance} and multi-GPU statevector simulation~\cite{brown2025multigpuquantumcircuitsimulation}, the present work addresses the complementary problem of executing independent circuits in parallel on both simulators and quantum hardware.

The remainder of this paper provides relevant background, describes the parallel execution methods and experimental setup, and presents results on both GPU simulators and quantum hardware. In the final sections, we discuss scaling, fidelity, and the future applicability of the techniques developed here.

\section{Background and Related Work}
\label{sec:background}

The parallel execution strategy introduced in this paper builds on several lines of prior work: recent proposals for multiprogramming quantum computers (MPQC), multi-GPU simulation infrastructure as a proxy for hardware quantum computers, and specific Pauli grouping techniques for observable estimation in Hamiltonian simulation.
We summarize these below.

\subsection{Prior Work on Hardware Parallel Execution}
\label{subsec:prior_hardware}

Parallel circuit execution on quantum hardware has two primary benefits. First, because noise limits the size of circuits that can be executed reliably, multiple smaller circuits can make better use of the available qubits. Second, combining multiple circuits into a single parallel submission can amortize the overhead associated with hardware initialization and data transfer, reducing execution time and cost.

Realizing these benefits is challenging, however. The compiler must identify partitions of the physical qubits onto which independent circuits can be mapped most efficiently. Concurrent gate operations introduce crosstalk that degrades fidelity, so partitioning must be crosstalk-aware. Co-located circuits may have unequal depths, and the resulting execution-time imbalance can introduce additional decoherence.

Quantum multi-programming compilers have been proposed to address these issues~\cite{das2019case, liu2021qucloud, niu2023enabling,Romo2026MultiQMN,Cao2025AnOR}.
Ohkura et al.\ characterize crosstalk during parallel execution and provide guidelines on inter-partition spacing~\cite{ohkura2022simultaneous,Niu2020AHH,Romo2026MultiQMN}.
Parallel execution has been shown to deliver substantial improvements for VQE~\cite{niu2022parallel,mineh2023accelerating,Zhao2026ASA} and Grover's search~\cite{park2023quantum}.

Our work builds upon one concrete implementation of MPQC, provided by Qiskit's \texttt{ParallelExperiment} class~\cite{qiskit_experiments}. This Qiskit library function provides the mechanism to compose multiple sub-experiments onto disjoint qubit sets and execute them as a single job, but leaves the important question of \emph{which qubits to assign} to the user, with no built-in mechanism for error-aware partition selection.
Our work addresses this gap by introducing a lightweight, topology-aware, partition mapping algorithm that accounts for one and two-qubit error data to achieve high-fidelity circuit execution with negligible computational overhead.

\subsection{Multi-GPU Simulation and Its Role as a Hardware Proxy}
\label{subsec:multigpu}

\begin{figure}
    \centering
    \includegraphics[width=1\linewidth]{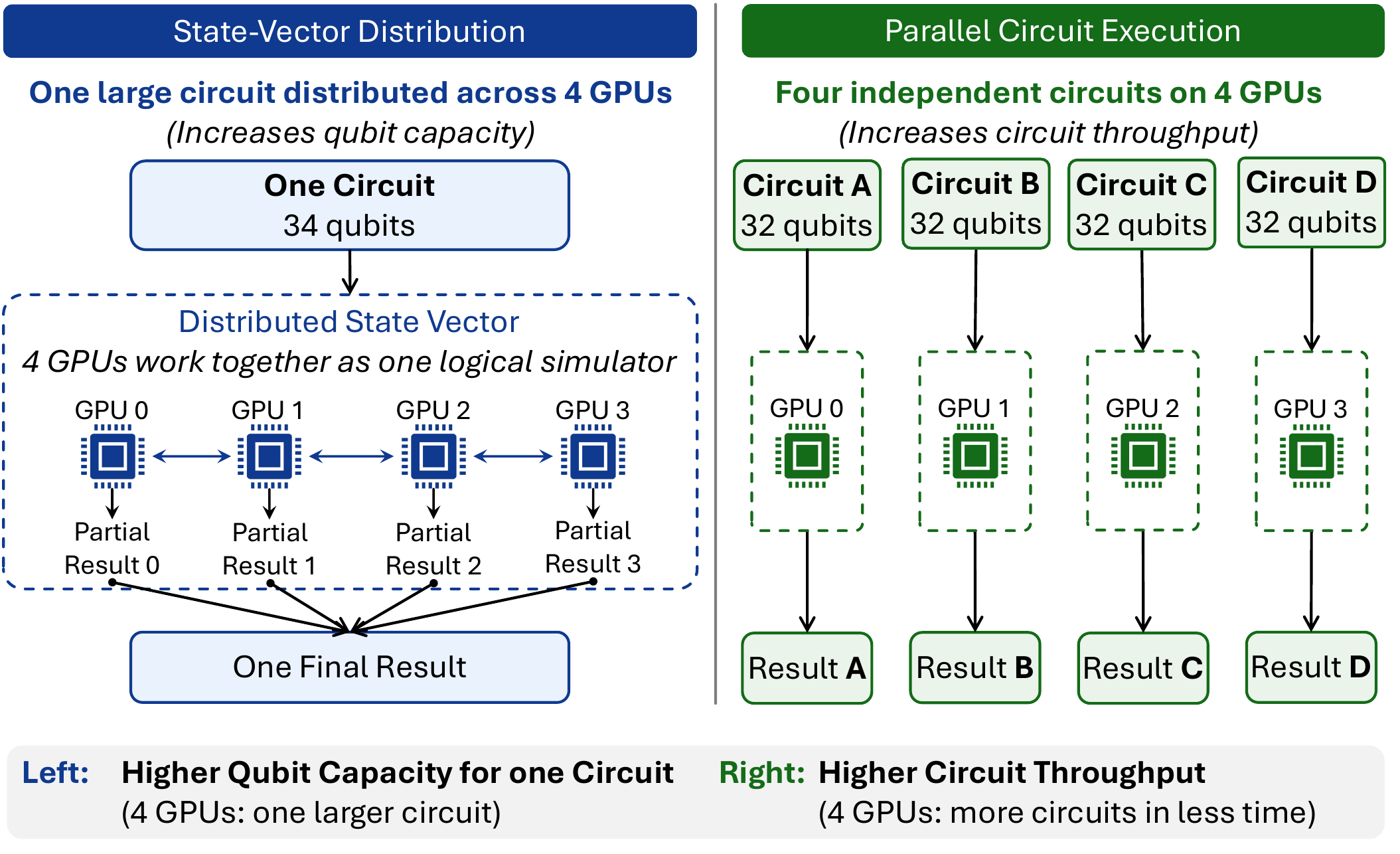}
   \caption{\textbf{Illustrative example of two multi-GPU execution paradigms.}
   The same four GPUs can either cooperate to simulate a single larger quantum circuit (left) or execute four independent circuits in parallel (right), trading increased qubit capacity for increased circuit throughput.}
\label{fig:state_vec_vs_par_exec}
\end{figure}

GPU statevector simulation is limited by the exponential storage requirement of $2^n$ complex amplitudes for $n$ qubits, restricting current systems to approximately 34 qubits on a single GPU and roughly 50 qubits on a large multi-GPU supercomputer. Within this range, simulation proceeds with 100\% fidelity, providing a useful environment for developing and validating execution strategies before deploying them on noisy quantum hardware.

NVIDIA's CUDA-Q framework~\cite{cudaq} provides several mechanisms for multi-GPU execution, including distributed statevector simulation and execution across multiple GPU-backed QPUs. As illustrated in~\autoref{fig:state_vec_vs_par_exec}, these represent two different forms of parallelism. In distributed statevector simulation (\texttt{mgpu}), multiple GPUs cooperate to simulate a single larger circuit by distributing its statevector across their combined memory. In circuit-level parallel execution, independent circuits execute on separate GPUs, increasing
circuit throughput rather than the maximum circuit width.

Brown et al.~\cite{brown2025multigpuquantumcircuitsimulation} examined multi-GPU statevector simulation in detail, introducing MPI support into the QED-C framework for HPC-scale execution. This approach is primarily suited to workloads with large individual circuits, where multiple GPUs increase the circuit size that can be simulated or reduce its execution time~\cite{chen2024multigpuenabledhybridquantumclassicalworkflow, xu2024atlashierarchicalpartitioningquantum}.

Circuit-level parallelism reduces wall-clock time for workloads containing many independent circuits. It also provides a useful proxy for the hardware parallel execution studied here: independent circuits are executed concurrently on separate GPU resources, analogous to their simultaneous execution on disjoint regions of a QPU. Because GPU simulation produces 100\% fidelity results, the same circuits also provide reference results for evaluating hardware fidelity and identifying errors introduced by routing, crosstalk, and other hardware effects.

Although statevector simulation can provide exact probabilities or expectation values directly, quantum hardware produces results through finite-shot measurement. For experiments intended to mirror hardware execution, we therefore use shot-based sampling on the GPU simulator. This preserves the hardware execution model while retaining the controlled, noise-free environment of simulation.

With sufficient independent circuits, execution across $N$ GPUs has a theoretical speedup of $N\times$, providing a performance ceiling against which the measured parallel execution can be compared. In practice, resource management, circuit distribution, and result collection introduce overhead that reduces the observed speedup.

\subsection{Observable Estimation and Hamiltonian Structure}
\label{subsec:observable}

A central computational task in variational algorithms such as VQE and in the estimation of observable values during Trotterized Hamiltonian simulation is the estimation of the expectation value of a Hamiltonian $H = \sum_i c_i P_i$, where $P_i$ are Pauli operators and $c_i$ are real coefficients. Execution on real quantum hardware requires the use of sampling, and each measurement circuit yields a statistical estimate of the expectation values of the Pauli operators it measures. Mutually commuting Pauli terms can be measured using a single circuit with appropriate basis rotations, reducing the number of distinct measurement circuits required.

The QED-C Application-Oriented Benchmark suite provides a framework for evaluating a wide variety of workloads across quantum hardware and simulation backends~\cite{QEDC-App-Benchmarks,Lubinski_2023}.  Recent extensions support Hamiltonian models from the HamLib library~\cite{sawaya2024hamliblibraryhamiltoniansquantum} and observable estimation using Pauli term grouping and optimized shot allocation~\cite{chatterjee2024comprehensivecrossmodelframeworkbenchmarking,niu2025practicalframeworkassessingperformance}. The present work enhances this framework by implementing functions to execute in parallel the independent measurement circuits produced by these Pauli groupings.

The number of commuting groups depends strongly on the Hamiltonian structure. A simple lattice model such as the Transverse-Field Ising Model (TFIM) decomposes into as few as 2 groups. A molecular electronic structure Hamiltonian such as H2 can produce hundreds to thousands of groups. If the circuits associated with these groups can be executed in parallel on independent computational resources, the potential speedup is limited only by the available resources and the number of independent measurement circuits.

\section{Methodology}
\label{sec:methods}

We describe our approach to parallel circuit execution across two quantum computing platforms: GPU-based simulators using the CUDA-Q~\cite{cudaq} SDK, and quantum hardware accessed using the Qiskit~\cite{qiskit_org} SDK. 
We use GPU simulation as a way to validate parallel execution methodology and the resulting performance metrics produced by the framework. We then introduce a practical implementation of parallel execution on quantum hardware devices using a qubit partitioning approach and discuss the tradeoffs associated with this method.

\subsection{Framework Overview}
\label{subsec:framework}

For practical application and scalability in quantum computation, parallel execution must be integrated into the execution pipeline of commonly used quantum algorithms.
To demonstrate this, our implementation is incorporated into the QED-C Application-Oriented Benchmarks framework~\cite{QEDC-App-Benchmarks,Lubinski_2023,lubinski2024quantumalgorithmexplorationusing,chatterjee2024comprehensivecrossmodelframeworkbenchmarking,niu2025practicalframeworkassessingperformance}, which provides a range of quantum programs from simple algorithms to complex applications which are run using a platform-agnostic circuit execution library (\emph{qedclib}) which includes built-in collection of performance metrics. 

Specifically, for our study, we make use of several of the algorithmic programs, e.g. Quantum Fourier Transform (QFT), Quantum Phase Estimation (QPE), and Hidden Shift (HS) to evaluate the impact of parallel execution. We execute an array of such circuits, both sequentially and in parallel, in order to quantify the reduction in execution time when circuits are executed in parallel as well as measure any reduction in the associated quality of result. 
Additionally, the benchmark suite provides an execution pipeline that accepts Hamiltonians from the HamLib library~\cite{sawaya2024hamliblibraryhamiltoniansquantum,chatterjee2024comprehensivecrossmodelframeworkbenchmarking}, performs Pauli grouping to generate measurement circuits, and executes those circuits using a configurable backend.

Our work adds a new parallel execution mode to this pipeline, selectable at runtime through a command-line argument, without modifying the circuit generation, grouping, or analysis components.  This allows us to measure the impact of practical parallel execution on both common algorithms and a realistic application in which the expectation value of an observable of a complex Hamiltonian can be quantified. 

\subsection{Multi-GPU Execution Modes}
\label{subsec:multi_gpu_modes}

NVIDIA's CUDA-Q framework~\cite{cudaq} provides several mechanisms for multi-GPU execution, including distributed statevector simulation and execution across multiple GPU-backed QPUs as described in~\autoref{subsec:multigpu}. These are fundamentally different ways of exploiting multiple GPUs for quantum circuit simulation. In our implementation, both use MPI-based coordination, they differ in what is distributed, how resources are allocated, and the type of workloads for which they are useful. 
In this section, we briefly describe our implementation of both within the \emph{qedclib} quantum circuit execution package.

\vspace{0.2cm}
\subsubsection{Statevector Distribution (\texttt{default} mode)}
\label{subsec:mgpu_paradigm}

In the first paradigm, all GPUs collaborate to simulate a single quantum circuit. This is the \texttt{qedclib} default multi-GPU mode and is automatically activated when execution is launched through \texttt{mpi4py} (using \texttt{-m mpi4py}, for example) and no other parallel options are specified. During CUDA-Q backend initialization, \texttt{mgpu} mode is specified. CUDA-Q then manages the distribution of the statevector across GPUs and returns the aggregated measurement results.
CUDA-Q partitions the state vector across GPU memories and uses inter-GPU communication when gate operations require access to amplitudes distributed across multiple GPUs~\cite{brown2025multigpuquantumcircuitsimulation, chen2024multigpuenabledhybridquantumclassicalworkflow, xu2024atlashierarchicalpartitioningquantum}.

This mode is primarily useful for simulating individual circuits that are too large for a single GPU or for reducing the execution time of large circuits. We include it here to distinguish this form of parallelism from the circuit-level parallel execution introduced in this work, in which independent circuits are distributed across GPU resources.

\vspace{0.2cm}
\subsubsection{Parallel Circuit Execution (\texttt{--parallel} mode)}
\label{subsec:parallel_paradigm}

In the second paradigm, independent circuits are distributed across available GPU resources, with no shared state between circuits. 
This mode is activated when execution is launched through \texttt{mpi4py} and the \texttt{--parallel} (or \texttt{-pm}) command line option is specified.

For GPU-based parallel execution, $T$ MPI ranks are divided into $G=T/P$ independent groups, where $P$ is the number of GPUs assigned to each circuit. Circuits are distributed in approximately equal blocks among the groups and executed independently. After execution, results are gathered in the designated leader rank and restored to their original circuit order before the standard benchmark post-processing and analysis are performed.

When $P = 1$, each circuit executes on a single GPU, providing circuit-level parallelism across the available GPUs, as illustrated in Fig.~\ref{fig:state_vec_vs_par_exec}. 

Our implementation uses MPI subcommunicators to create independent GPU groups. This provides fine-grained control of the MPI implementation for finite-shot sampling.
For a workload containing $C$ independent circuits, the theoretical maximum speedup is $\min(G,C)$, where $G=T/P$ is the number of independent execution groups. When $C \gg G$, all groups can remain active and near-linear circuit-level scaling is possible. When $C < G$, the available circuit-level parallelism is limited by $C$.

\vspace{0.2cm}
\subsubsection{Hybrid Parallel Execution (\texttt{--gpus\_per\_circuit} mode)}
\label{subsec:combined}

When $P > 1$, multiple GPUs cooperate to simulate each circuit using the statevector distribution method described in Section III-B1, while multiple circuits execute concurrently. This combines statevector-level and circuit-level parallelism.

In this hybrid approach, each of $G_1$ circuit groups is assigned to a block of $G_2$ GPUs. Each block uses \texttt{mgpu} mode internally to simulate its assigned circuit, while the blocks operate independently in parallel.
The \texttt{--gpus\_per\_circuit} (or \texttt{-gpc}) option specifies the number of GPUs assigned to each circuit. A value greater than one activates this hybrid mode of execution.

For a total of $G = G_1 \times G_2$ GPUs, this enables simultaneous parallelism at both the circuit level and the statevector level, benefiting both from extended qubit capacity and from reduced sequential circuit execution.

\vspace{0.3cm}

Overall, our implementation provides a higher-level parallel execution capability in which a workload containing independent circuits can be executed across multiple GPUs without requiring the application to manage their distribution or the collection of results. 
The same workload can be executed sequentially, using circuit-level parallelism, or using hybrid circuit- and statevector-level parallelism without changing the logic used to generate the circuits or analyze the results.
Additional details of this implementation can be found in Appendix~\ref{app:multi_gpu_details}.


\begin{figure*}
    \centering
    \includegraphics[width=0.9\linewidth]{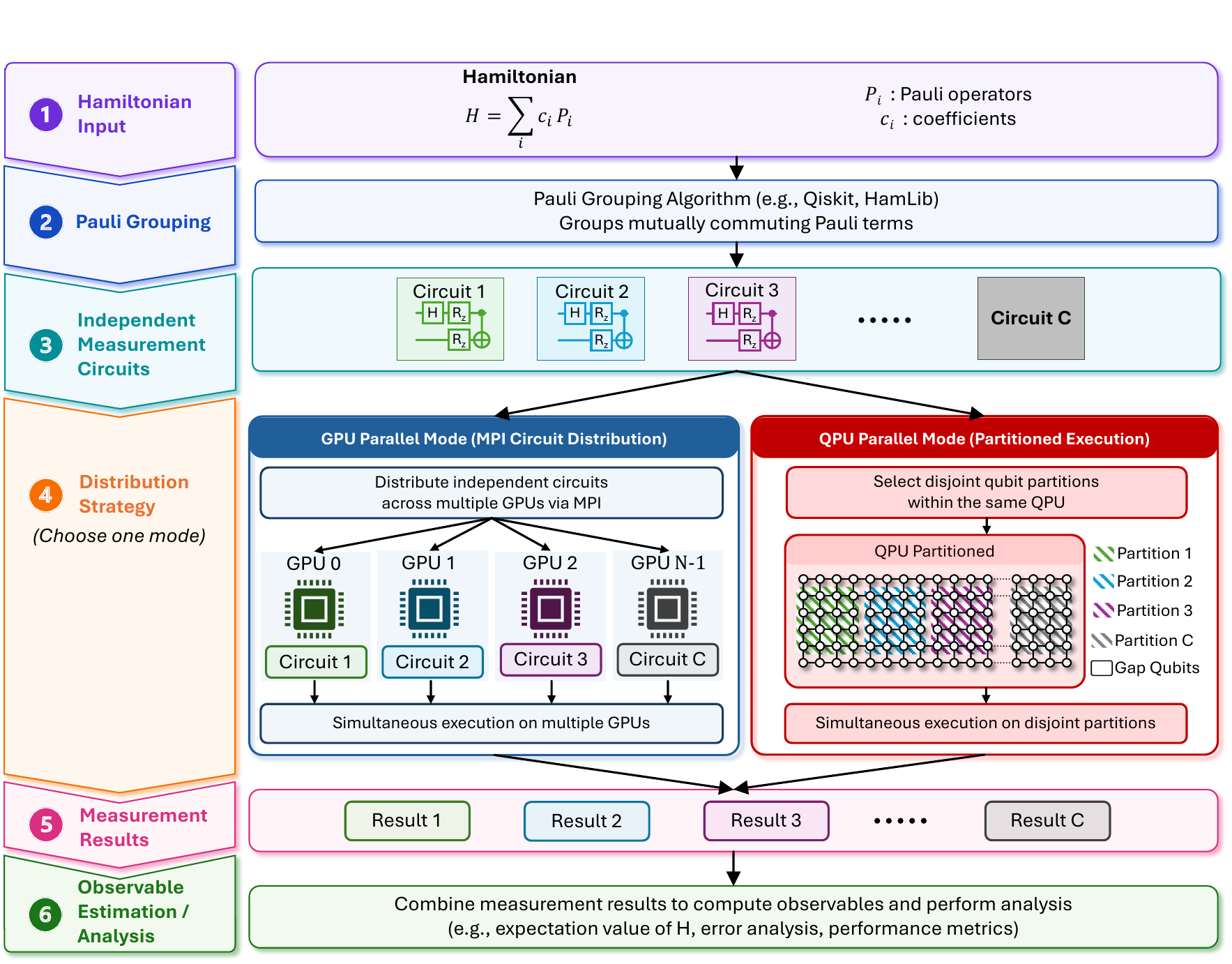}
   \caption{\textbf{Parallel Execution Workflow for Observable Estimation.}
    A Hamiltonian is first decomposed into commuting Pauli groups, producing a collection of independent measurement circuits. These circuits are then executed using one of two backends: (i) distributed across multiple GPUs using MPI-based circuit-level parallelism, or (ii) mapped onto disjoint qubit partitions within a single QPU for simultaneous hardware execution. Each circuit produces an independent measurement result, which is subsequently used for observable estimation and benchmark analysis. This workflow forms the basis for both the GPU simulation and hardware execution strategies presented in this paper.}
\label{fig:exec_flow}
\end{figure*}


\begin{table*}[h]
\centering
\caption{Overview of the benchmark workloads used to evaluate the parallel framework on GPU and QPU platforms.}
\label{tab:gpu-qpu-summary}
\setlength{\tabcolsep}{3.7pt}

\begin{tabular}{l|l|cccc|cccc}
\toprule
\multicolumn{1}{c|}{\multirow{2}{*}{\textbf{Type}}}
&
\multicolumn{1}{c|}{\multirow{2}{*}{\textbf{Benchmark}}}
&
\multicolumn{4}{c|}{\textbf{GPU Experiments}}
&
\multicolumn{4}{c}{\textbf{QPU Experiments}}
\\
\cmidrule(lr){3-6}
\cmidrule(l){7-10}

\multicolumn{1}{c|}{}
&
\multicolumn{1}{c|}{}
&
\textbf{\# Qubits}
&
\textbf{Circuit Depth}
&
\textbf{\# Terms}
&
\textbf{\# Groups}
&
\textbf{\# Qubits}
&
\textbf{Circuit Depth}
&
\textbf{\# Terms}
&
\textbf{\# Groups}
\\
\midrule

\multirow{5}{*}{%
\begin{tabular}[c]{@{}l@{}}
Standard QED-C\\
Benchmark Experiments
\end{tabular}}
& Quantum Fourier Transform
& 25--41 & 2500--6724 & -- & --
& 5--16 & 23--275 & -- & -- \\

& Phase Estimation
& 25--41 & 2500--6724 & -- & --
& -- & -- & -- & -- \\

& Hidden Shift
& 26--40 & 2704--6400 & -- & --
& -- & -- & -- & -- \\

& Bernstein--Vazirani
& -- & -- & -- & --
& 5--16 & 6--16 & -- & -- \\

& Hamiltonian Sim. (TFIM)
& -- & -- & -- & --
& 5--16 & 40--103 & -- & -- \\

\midrule

\multirow{3}{*}{%
\begin{tabular}[c]{@{}l@{}}
Observable Estimation\\
Experiments
\end{tabular}}
& TFIM
& 20--28 & -- & 40--56 & 2
& 5--16 & -- & 8--32 & 2 \\

& Bose-Hubbard
& 20--28 & -- & 319--459 & 9--11
& 6--16 & -- & 39--249 & 9--12 \\

& H2
& 8--20 & -- & 185--2951 & 54--1251
& 6--12 & -- & 15--327 & 3--59 \\

\bottomrule
\end{tabular}
\end{table*}

\subsection{Parallel Execution on Quantum Hardware}
\label{subsec:hardware_implementation}

The parallel circuit execution strategy described in this work is designed with an eye toward an analogous implementation on real quantum hardware.
On quantum hardware, independent circuits can be executed concurrently on separate QPUs or on disjoint qubit regions of a single large QPU. The latter approach is the focus of the hardware implementation in this work.

Parallel execution on a partitioned QPU requires addressing three
hardware-specific problems: (1) selecting disjoint qubit regions with
low error rates, (2) mapping circuits onto those regions, and
(3) assembling the mapped circuits into a single job submission.

To address these challenges, we make use of 
the open source classes \texttt{ParallelExperiment} and \texttt{BaseExperiment} from the \texttt{qiskit-experiments} package~\cite{qiskit_experiments}. These provide the circuit composition and result decomposition infrastructure required for our work.  We defined our own custom partitioning and orchestration logic to make use of the capabilities of these classes. 

For execution on a partitioned QPU, independent circuits are mapped to disjoint regions of the device and combined into a composite circuit for simultaneous execution. Given $N_p$ available partitions, up to $N_p$ circuits can be executed concurrently. Following execution, the measurement results are separated to recover the outcomes of the individual circuits. In the ideal case, this provides a speedup of up to $N_p$ over sequential execution, subject to the quality and connectivity of the available partitions and the requirement that simultaneous execution does not significantly degrade result quality.

\vspace{0.3cm}
\subsubsection{Partition Finding}

We use a lightweight topology- and error-aware algorithm to identify disjoint qubit regions for the required circuit widths.
Starting from each qubit on the device, we construct a connected subgraph of the target width using greedy frontier expansion. The algorithm selects nodes with the most connections to the existing cluster, favoring compact, well-connected regions over chains.
 
Each candidate partition is scored by three criteria in priority order:
\begin{enumerate}[label=\alph*.]
    \item {Average 2-qubit gate error:} Obtained from the backend calibration data (\texttt{backend.target}). Partitions in higher-error regions of the device are ranked lower.
    \item {Diameter:} The maximum shortest path between any two qubits in the partition. Lower diameter indicates a more compact subgraph with shorter routing paths.
    \item {Internal edge count:} More edges within the partition provide the transpiler with greater routing flexibility.
\end{enumerate}

Partitions are selected greedily in score order, with a configurable \emph{gap} parameter (default: 2 hops) enforcing minimum distance between any two partitions' qubits.
This gap provides additional separation between concurrently executing circuits at the cost of reduced partition count.
If insufficient partitions are found at the default gap, the algorithm retries with progressively smaller gaps (2 $\to$ 1 $\to$ 0). Additionally, partitions whose average 2-qubit gate error exceeds a threshold (default: 0.02) are rejected entirely, preventing circuits from landing on degraded device regions.

The algorithm is not specific to IBM hardware, as it requires only the device coupling map and gate calibration data, although our experiments were performed only on IBM devices. Details of the algorithms used for partition-finding, scoring, and selection are provided in Appendix~\ref{app:partition_mapping}.

\vspace{0.3cm}
\subsubsection{Circuit Distribution and Array Batching}

Given $N$ circuits and $P$ partitions, circuits are distributed round-robin across partitions, producing arrays of $\lceil N/P \rceil$ circuits per partition.
We leverage the \texttt{ParallelExperiment} framework's native support for array experiments: each partition is represented as a \texttt{CircuitArrayExperiment} that returns multiple circuits, and \texttt{ParallelExperiment} composes them by index into $\lceil N/P \rceil$ composite circuits, all submitted as a \emph{single job}.

This array batching approach reduces the number of job submissions from $N$ (one per circuit) or $\lceil N/P \rceil$ (one per round) to exactly 1, eliminating repeated queue wait times and job initialization overhead.

\vspace{0.3cm}
\subsubsection{Mixed-Width Handling}

Some applications produce circuits of varying widths (e.g., a sweep from 2 to 20 qubits). We address this using a two-phase partition allocation:
\begin{enumerate}[label=\alph*.]
    \item {Phase 1:} Allocate one partition per unique circuit width, starting from the largest. Stop when the device cannot accommodate the next width.
    \item {Phase 2:} Allocate additional partitions for widths already represented, reducing the number of rounds.
\end{enumerate}

Circuits matching a partition's width are assigned directly. Remaining circuits are padded with idle qubits and distributed round-robin across all available partitions to balance the load.
The result is a single \texttt{ParallelExperiment} containing partitions of different sizes, all submitted as one job.

\vspace{0.3cm}
\subsubsection{Transpiler Integration}

A subtle challenge stems from the interaction between parallel execution and circuit transpilation. When \texttt{ParallelExperiment} transpiles its constituent circuits, each sub-experiment is assigned a set of \texttt{physical\_qubits} on the device. For shallow circuits, the transpiler respects these boundaries. For deeper circuits, however, the transpiler's routing algorithm may insert SWAP gates that route through qubits outside the assigned partition, causing a conflict.
We handle this with a try/fallback strategy:
\begin{enumerate}[label=\alph*.]
    \item First, run with full transpiler optimization (level 1), which gives best fidelity when the circuit stays within bounds.
    \item If the transpiler escapes the partition, fall back to \emph{pre-transpilation}: construct a restricted coupling map from only the edges within each partition, transpile each circuit onto this restricted map, then run \texttt{ParallelExperiment} with optimization level 0 (no re-transpilation).
\end{enumerate}

The pre-transpile fallback guarantees partition containment at a cost of 10--25\% fidelity reduction due to additional SWAP insertions in the constrained topology.

\subsection{Parallel Execution in Observable Estimation}
\label{subsec:observable_execution}

Observable estimation is particularly well suited to parallel circuit execution because Pauli grouping produces a set of measurement circuits that can be executed independently. Once the Hamiltonian has been decomposed and the measurement circuits constructed, the execution stage can use the same parallel mechanisms described above. The circuits can be distributed across GPU resources or assigned to disjoint partitions of a QPU, after which the measurement results are returned to the existing observable-analysis pipeline.

Figure~\ref{fig:exec_flow} shows the separation between the generation of independent measurement circuits, their parallel execution, and the subsequent observable computation. This separation allows the same observable-estimation workflow to be used with either GPU simulation or quantum hardware, while changing only the manner in which the measurement circuits are executed.

For Hamiltonian simulation workloads, the measurement circuits can be distributed at either the circuit or measurement-group level. In circuit-level execution, individual measurement circuits are distributed independently across the available computational resources and use a common shot count. This is appropriate when each circuit can be treated as an independent unit of work, as is the case for the measurement circuits produced by Pauli grouping.

For observable calculations using optimized shot allocation, the measurement group provides a more useful unit of distribution. Circuits belonging to the same measurement group are kept together, while different groups can be executed with independently assigned shot counts. This avoids separating circuits that share a common shot-allocation requirement and preserves the allocation determined by the observable-estimation procedure.

The same decomposition provides a common workload for comparing parallel execution on GPUs and quantum hardware. On GPUs, independent measurement circuits or groups are assigned to separate GPU resources through the MPI execution path. On quantum hardware, the corresponding circuits are assigned to disjoint qubit partitions and executed concurrently. In both cases, the parallel execution layer changes how the measurement workload is scheduled, while the Hamiltonian decomposition, Pauli grouping, and observable analysis remain unchanged.


\begin{table*}[h]
\centering
\caption{Average speedups measured across the GPU and QPU experiments. Standard QED-C and observable-estimation GPU results use circuit-level parallel execution, while QPU results use partition-based parallel circuit execution. GPU speedups are computed relative to the 1-GPU baseline and averaged over all evaluated problem sizes. Statevector-distribution results are reported separately in Appendix~D.}
\label{tab:benchmark_summary_results}
\setlength{\tabcolsep}{3.7pt}
\begin{tabular}{@{}l|l|ccccc|cc@{}}
\toprule

\multicolumn{1}{c|}{\multirow{2}{*}{\textbf{Type}}} &
\multicolumn{1}{c|}{\multirow{2}{*}{\textbf{Benchmark}}} &
\multicolumn{5}{c|}{\textbf{GPU Experiments}} &
\multicolumn{2}{c}{\textbf{QPU Experiments}} \\

\cmidrule(lr){3-7}
\cmidrule(l){8-9}

& &
\textbf{\# Qubits} &
\textbf{Avg. Speedup} &
\textbf{Avg. Speedup} &
\textbf{Avg. Speedup} &
\textbf{Overall Avg.} &
\textbf{\# Qubits} &
\textbf{Avg. Speedup} \\

& &
&
\textbf{(4 GPU)} &
\textbf{(8 GPU)} &
\textbf{(16 GPU)} &
\textbf{Speedup} &
&
\\

\midrule

\multirow{5}{*}{Standard QED-C}
& Quantum Fourier Transform
& 25--41
& 1.71$\times$
& 1.37$\times$
& 2.12$\times$
& 1.73$\times$
& 5--16
& 3.64$\times$ \\

& Phase Estimation
& 25--41
& 1.99$\times$
& 1.35$\times$
& 1.95$\times$
& 1.76$\times$
& --
& -- \\

& Hidden Shift
& 26--40
& 1.38$\times$
& 0.76$\times$
& 1.08$\times$
& 1.07$\times$
& --
& -- \\

& Bernstein--Vazirani
& --
& --
& --
& --
& --
& 5--16
& 3.58$\times$ \\

& Hamiltonian Sim. (TFIM)
& --
& --
& --
& --
& --
& 5--16
& 3.33$\times$ \\

\multicolumn{2}{l|}{\cellcolor{summarygray}\textbf{Standard QED-C Average}}
& \cellcolor{summarygray}--
& \cellcolor{summarygray}\textbf{1.69$\times$}
& \cellcolor{summarygray}\textbf{1.16$\times$}
& \cellcolor{summarygray}\textbf{1.72$\times$}
& \cellcolor{summarygray}\textbf{\textcolor{BrickRed}{1.52$\times$}}
& \cellcolor{summarygray}--
& \cellcolor{summarygray}\textbf{\textcolor{BrickRed}{3.52$\times$}} \\

\midrule

\multirow{3}{*}{Observable Estimation}
& TFIM
& 20--28
& 0.81$\times$
& 0.79$\times$
& 1.53$\times$
& 1.04$\times$
& 5--16
& 4.44$\times$ \\

& Bose-Hubbard
& 20--28
& 2.19$\times$
& 2.90$\times$
& 5.88$\times$
& 3.66$\times$
& 6--16
& 8.68$\times$ \\

& H2
& 8--20
& 3.05$\times$
& 4.90$\times$
& 9.96$\times$
& 5.97$\times$
& 6--12
& 2.48$\times$ \\

\multicolumn{2}{l|}{\cellcolor{summarygray}\textbf{Observable Estimation Average}}
& \cellcolor{summarygray}--
& \cellcolor{summarygray}\textbf{2.02$\times$}
& \cellcolor{summarygray}\textbf{2.86$\times$}
& \cellcolor{summarygray}\textbf{5.79$\times$}
& \cellcolor{summarygray}\textbf{\textcolor{BrickRed}{3.56$\times$}}
& \cellcolor{summarygray}--
& \cellcolor{summarygray}\textbf{\textcolor{BrickRed}{5.20$\times$}} \\

\bottomrule
\end{tabular}

\end{table*}

\begin{figure*}
  \centering
  \includegraphics[width=1\textwidth]{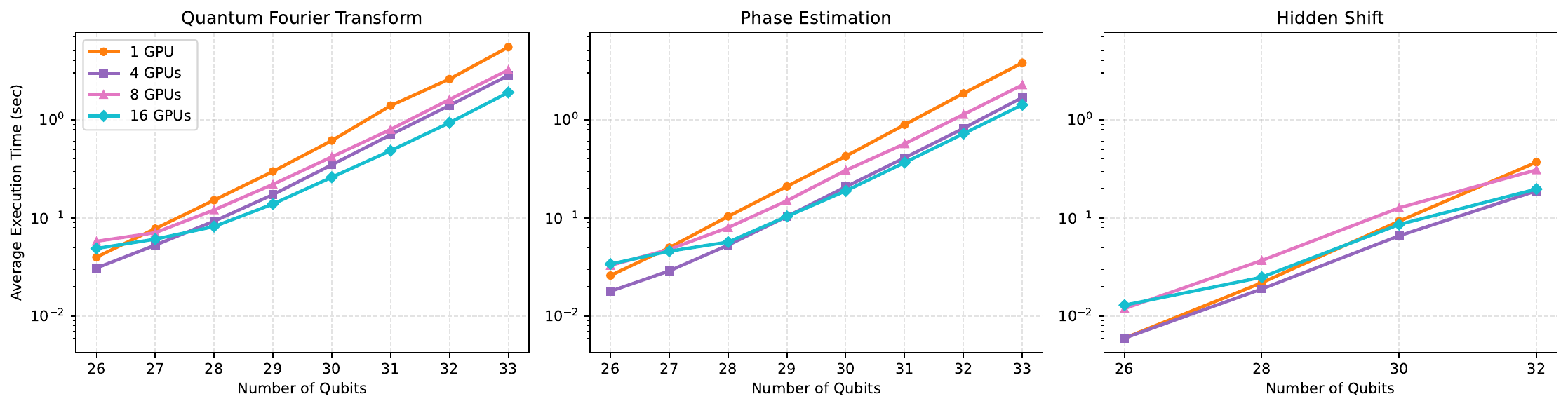}
    \par\medskip
    \includegraphics[width=1\textwidth]{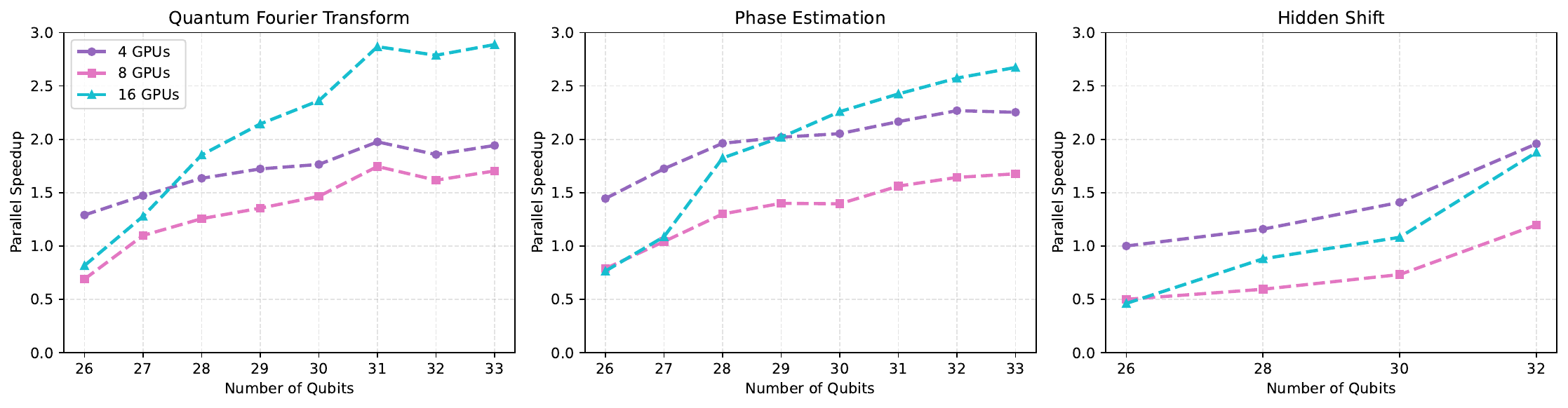}
    \caption{\textbf{GPU Scaling for Standard QED-C Benchmark Applications.}
    Average execution time (top row) and parallel speedup relative to a single GPU (bottom row) for the Quantum Fourier Transform (QFT), Quantum Phase Estimation (QPE), and Hidden Shift (HS) benchmarks executed on NVIDIA Perlmutter GPUs using 1, 4, 8, and 16 GPUs. The top row reports the average execution time as a function of qubit count on a logarithmic scale, while the bottom row shows the corresponding parallel speedup achieved with 4, 8, and 16 GPUs relative to single-GPU execution. Scaling is non-monotonic: performance can degrade from 4 to 8 GPUs as execution extends beyond a single NVLink-connected node and incurs inter-node communication overhead, while at 16 GPUs the additional parallel resources increasingly offset this cost for larger problem sizes. Because these experiments use noiseless GPU-based statevector simulation, the output fidelity is effectively 100\% for all configurations; therefore, separate fidelity plots are omitted.}
  \label{fig:gpu_benchmarks}
\end{figure*}

\section{Experimental Setup}
\label{sec:setup}

\subsection{Hardware Environment}

All GPU-based experiments reported in this work were conducted on the NERSC Perlmutter system~\cite{Perlmutter} at Lawrence Berkeley National Laboratory. Perlmutter is a heterogeneous HPC system whose GPU-accelerated partition consists of nodes each equipped with four NVIDIA A100 40 GB GPUs connected by NVLink, with nodes interconnected via the HPE Slingshot 11 high-speed network fabric. Experiments using up to 16 GPUs spanned 4 nodes (4 GPUs per node), and experiments at larger GPU counts used proportionally more nodes. 

Hardware experiments were performed on IBM's \texttt{ibm\_boston} processor (156 qubits, Heron r3 architecture) accessed via the IBM Quantum Network through NERSC. 

\subsection{Software Environment}

The QED-C Application-Oriented Benchmark suite was installed in developer mode from the \texttt{QC-App-Oriented-Benchmarks} repository~\cite{QEDC-App-Benchmarks}.
GPU experiments used CUDA-Q version 0.13.0 as the quantum simulation backend, with \texttt{mpi4py} for MPI coordination and Python 3.12.
CUDA-Q version 0.15 added support for MPI subcommunicators to allow greater flexibility in the partitioning of HPC resources for simultaneous execution with multiple circuits and/or QPUs. In this work, we used this extended API, backported to CUDA-Q 0.13, to implement support for hybrid parallelism of simulations in the QED-C \texttt{qedclib} package.

Hardware QPU experiments used Qiskit 2.5 with qiskit-ibm-runtime 0.48.0 and qiskit-experiments 0.14.0 for circuit composition and result decomposition.

\subsection{Standard QED-C Benchmarks}

We evaluate our framework using standard application-oriented benchmarks from the QED-C benchmark suite, including Bernstein--Vazirani (BV), the Quantum Fourier Transform (QFT), Quantum Phase Estimation (QPE), Hidden Shift (HS), and the Transverse Field Ising Model (TFIM). Table~\ref{tab:gpu-qpu-summary} summarizes the experimental workloads evaluated in this study, including the benchmark suites, problem sizes, and workload characteristics on GPU and QPU platforms.

These benchmarks are used to assess the framework on both GPU simulators and IBM quantum hardware. On GPUs, QFT, QPE, and Hidden Shift are evaluated using circuit-level parallel execution across 1, 4, 8, and 16 GPUs, as reported in Section~\ref{sec:results}. The same benchmarks are also evaluated separately using \texttt{mgpu} statevector distribution across up to 256 GPUs, with those results reported in Appendix~D. On QPUs, BV, QFT, and TFIM are used to evaluate the impact of partition-based parallel execution on execution time, fidelity, and resource utilization.

Observable estimation experiments used three Hamiltonians extracted from the HamLib library via the QED-C \texttt{qedcbench} module on both GPU simulators and IBM quantum hardware:

\begin{enumerate}[label=\alph*.]

\item {TFIM}: 1D Transverse-Field Ising Model with periodic boundary conditions ($h=2$). Qubit widths: 20--28. Number of Pauli terms: 40--56. Number of commuting groups (circuits): {2}. \texttt{condensedmatter/tfim/tfim}

\item {Bose-Hubbard}: 1D Bose-Hubbard model with gray encoding ($U=10$). Qubit widths: 20--28. Number of Pauli terms: 319--459. Number of commuting groups (circuits): {9--11}. For hybrid parallelism, qubit width: 36. Number of commuting groups (circuits): {24}. \texttt{condensedmatter/bosehubbard/BH\_D-1\_d-4}

\item {H2}: Hydrogen molecule electronic structure with Bravyi--Kitaev encoding. Qubit widths: 8, 12, 20. Number of Pauli terms: 185--2951. Number of commuting groups (circuits): {54--1251}. \texttt{chemistry/electronic/standard/H2}

\end{enumerate}

These benchmarks were selected to span a broad range of observable estimation complexity, ranging from two commuting measurement groups for the TFIM Hamiltonian to more than one thousand groups for the H2 molecular Hamiltonian. For the hybrid parallel execution experiments, an additional 36-qubit Bose-Hubbard instance containing 24 commuting groups was also evaluated. Unless stated otherwise, HamLib experiments used $K=1$ Trotter step, time $t=0.1$, and 10,000 shots per circuit. Each data point represents the execution time for a single benchmark run; reported values are consistent across multiple runs.

\section{Results}
\label{sec:results}

This section evaluates parallel circuit execution on GPU simulators and IBM quantum hardware. GPU experiments quantify the scalability and performance of circuit-level parallel execution in a noiseless environment. Hardware experiments evaluate the practical execution-time reduction and fidelity tradeoffs achievable on current quantum processors. Table~\ref{tab:benchmark_summary_results} summarizes the benchmark workloads considered in this evaluation and the average speedups achieved by the parallel execution framework across all evaluated problem sizes.


\subsection{Evaluation on GPU Simulators}

We first present results obtained from execution on GPU simulators using CUDA-Q on the NERSC Perlmutter system.

\vspace{0.3cm}
\subsubsection{Standard QED-C Benchmark Algorithms}

The Quantum Fourier Transform (QFT), Quantum Phase Estimation (QPE), and Hidden Shift (HS) programs were executed using 1, 4, 8, and 16 GPUs.  Fig.~\ref{fig:gpu_benchmarks} presents the average execution time and the corresponding parallel speedup for each of these benchmark algorithms.

Fig.~\ref{fig:gpu_benchmarks} (top) shows that execution time increases with circuit size for all three benchmarks, while distributing the execution across multiple GPUs consistently reduces the runtime. The reduction is more pronounced as the number of qubits grows, indicating that the additional parallel resources are more effectively utilized for larger problem instances.

The corresponding parallel speedups relative to single-GPU execution are shown in Fig.~\ref{fig:gpu_benchmarks} (bottom). For QFT and QPE, the speedup increases steadily with circuit size, demonstrating improved scalability as the computational workload grows. Hidden Shift exhibits more modest gains for smaller circuits, where communication overhead represents a larger fraction of the total runtime, but the achieved speedup also improves for larger problem sizes. Overall, the results demonstrate that this implementation effectively exploits multiple GPUs and provides increasing performance benefits as circuit complexity increases.


\begin{figure*}[]
  \centering
  \includegraphics[width=1\textwidth]{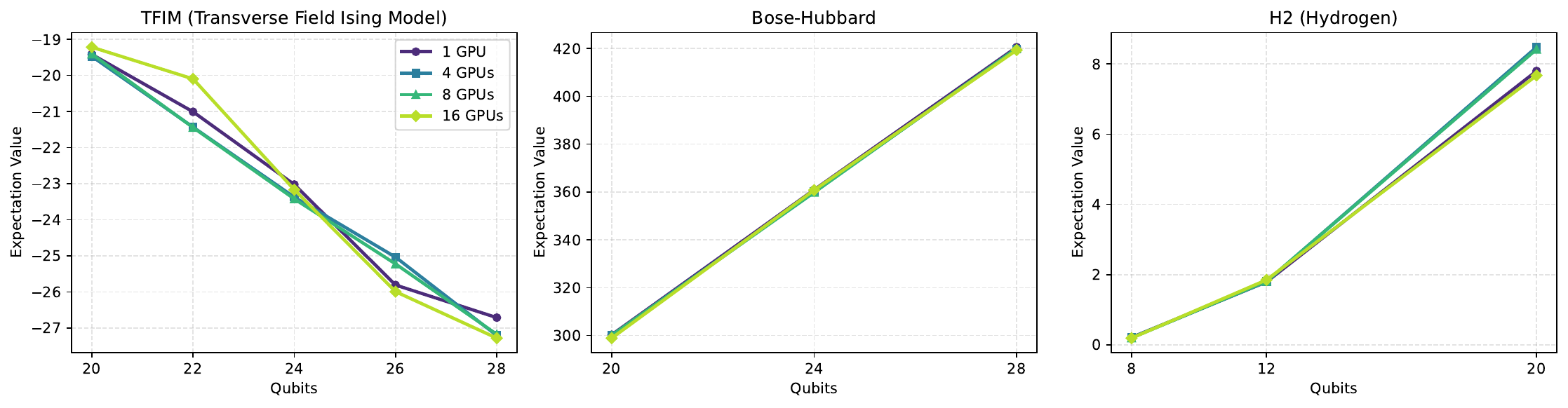}
    \par\medskip
    \includegraphics[width=1\textwidth]{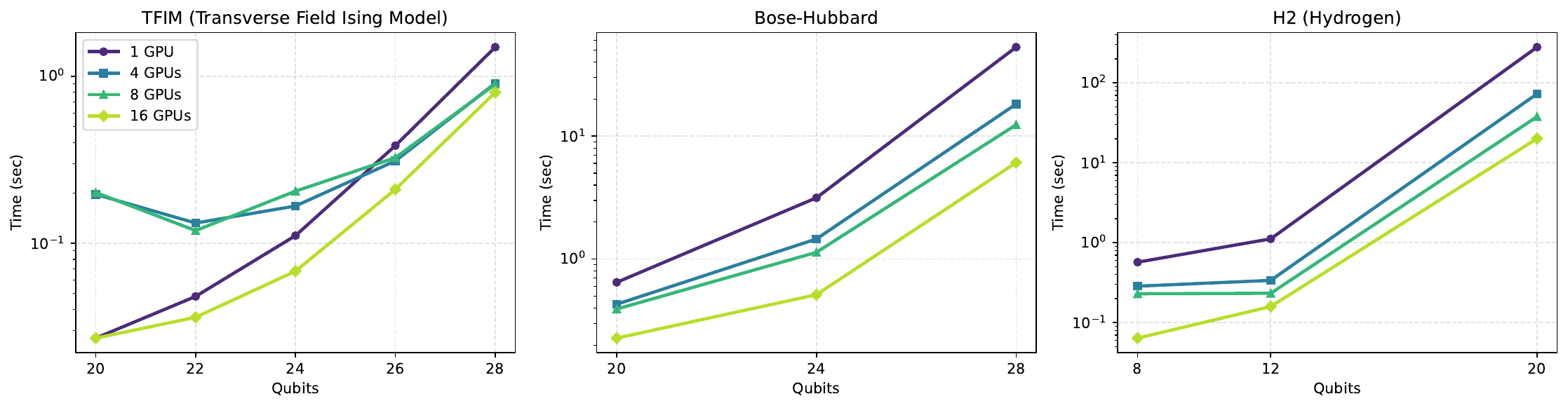}
    \caption{\textbf{GPU Observable-Estimation Consistency and Execution-Time Scaling.} Sampled expectation values (top row) and execution times (bottom row) for the transverse-field Ising model (TFIM), Bose–Hubbard model, and $H_2$ electronic-structure Hamiltonian executed on NERSC Perlmutter using 1, 4, 8, and 16 NVIDIA A100 GPUs. The top row compares the expectation values obtained through shot-based sampling across the different GPU configurations. Results remain closely aligned across GPU counts, demonstrating that distributing independent measurement circuits across GPUs does not introduce a systematic change in the sampled expectation values. The bottom row reports execution time on a logarithmic scale. TFIM contains only two commuting groups and therefore shows limited improvement beyond two GPUs, Bose–Hubbard contains 9–11 groups and scales effectively across intermediate GPU counts, and $H_2$ contains up to 1,251 groups and provides sufficient independent circuits to utilize all 16 GPUs.}
  \label{fig:gpu_scaling}
\end{figure*}

\begin{table}
\centering
\caption{Performance summary for GPU-based observable estimation at the largest qubit width evaluated for each Hamiltonian using 16 NVIDIA A100 GPUs on NERSC Perlmutter.}
\label{tab:gpu-observable-results}
\setlength{\tabcolsep}{5pt}
\begin{tabular}{lccccc}
\toprule
\textbf{Hamiltonian} &
\textbf{\# Qubits} &
\textbf{\# Groups} &
\textbf{1 GPU} &
\textbf{16 GPUs} &
\textbf{Speedup} \\
\midrule
TFIM           & 28 & 2    & 1.70   & 0.90  & $1.9\times$ \\
Bose-Hubbard  & 28 & 9    & 53.54  & 6.50  & $8.2\times$ \\
H2          & 20 & 1251 & 283.81 & 20.51 & $13.8\times$ \\
\bottomrule
\end{tabular}
\end{table}


\begin{figure*}
    \includegraphics[width=1\textwidth]{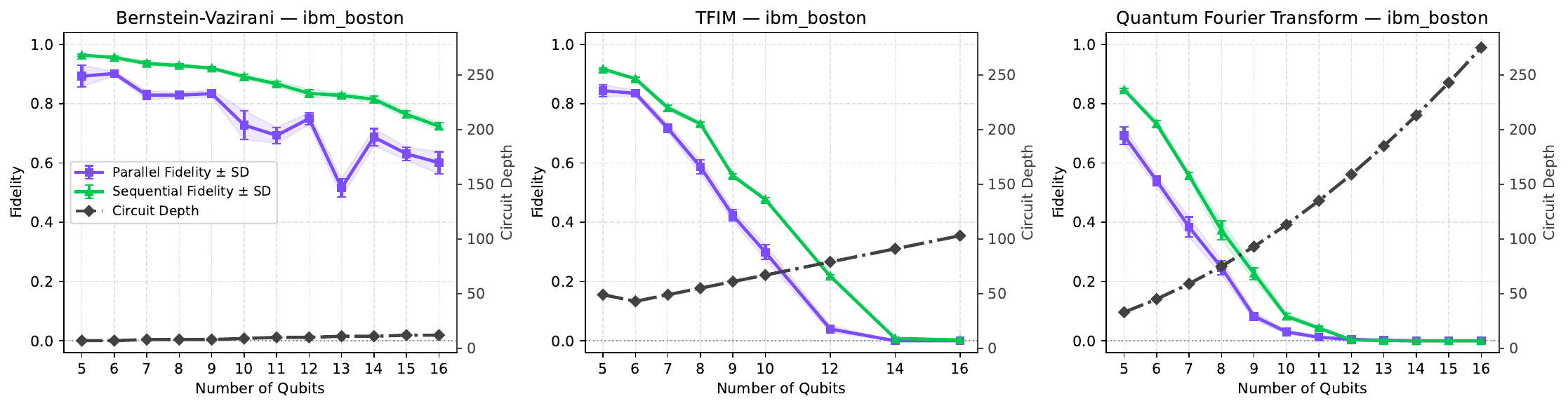}
    \par\medskip
    \includegraphics[width=1\textwidth]{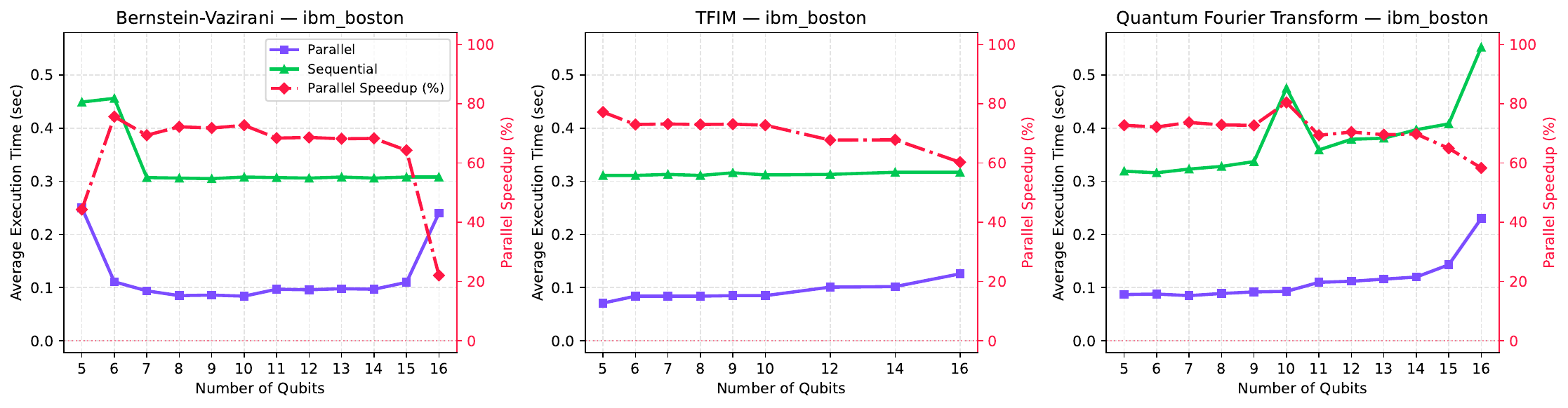}
    \caption{\textbf{Fidelity and Billed Execution-Time Comparison for Standard QED-C Benchmarks on IBM Boston.}
    Parallel and sequential execution results for the Bernstein--Vazirani (BV), transverse-field Ising model (TFIM), and Quantum Fourier Transform (QFT) benchmarks. The top row reports the Hellinger fidelity as a function of qubit width, with error bars indicating the standard deviation across repeated runs; the secondary axis shows the corresponding transpiled circuit depth. Parallel execution generally exhibits lower fidelity than sequential execution because partitioning constrains qubit placement and routing within each partition, introducing additional routing overhead. As circuit depth increases, particularly for TFIM and QFT, the fidelity of both execution strategies decreases, with the parallel implementation experiencing a larger degradation. The bottom row reports the billed execution time together with the percentage reduction achieved through parallel execution. Executing independent circuits simultaneously on disjoint qubit partitions consistently reduces billed execution time, although the achievable reduction decreases for larger circuits as the number of available partitions becomes more limited.}
    \label{fig:fidelity_and_time_boston}
\end{figure*}

\begin{figure*}[]
    \includegraphics[width=1\textwidth]{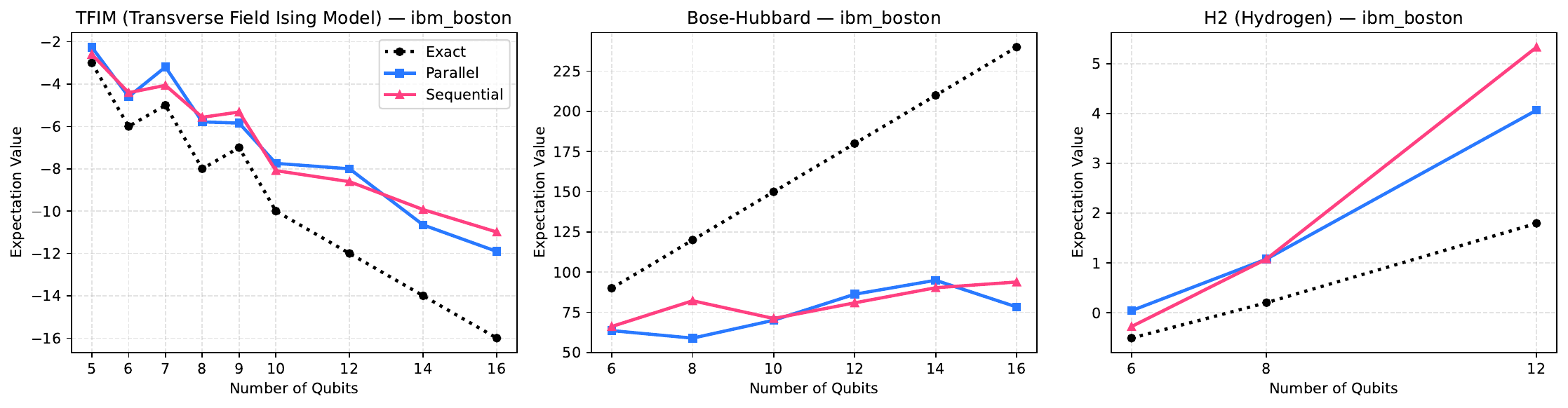}
    \par\medskip
    \includegraphics[width=1\textwidth]{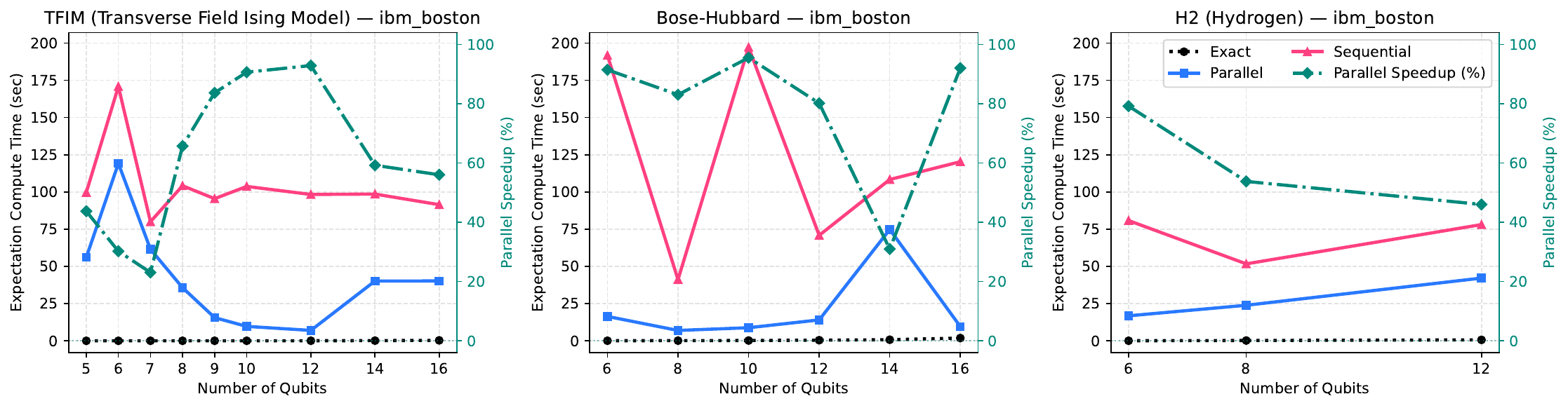}
    \caption{\textbf{Expectation value accuracy and computational performance of sequential and parallel execution.} Comparison of the exact, parallel, and sequential expectation values for the H2, Bose-Hubbard, and Transverse Field Ising Model (TFIM) Hamiltonians executed on the IBM Boston backend (top row). The bottom row compares the corresponding expectation value computation times, together with the parallel speedup relative to sequential execution (right-hand axis). Positive speedup values indicate reduced computation time achieved by the parallel implementation.}
    \label{fig:hamlib_expectation_and_time_boston}
\end{figure*}

\vspace{0.3cm}
\subsubsection{Observable Estimation}

We then evaluated GPU-based parallel circuit execution for estimating observables using 1, 4, 8, and 16 GPUs for three Hamiltonians: the transverse-field Ising model (TFIM), the Bose-Hubbard model (BH), and the H2 electronic-structure Hamiltonian (H2). All experiments used the simple sampling method, in which commuting Pauli terms are grouped into independent measurement circuits and executed using shot-based sampling. Unless otherwise stated, the experiments used $K=1$ Trotter step, evolution time $t=0.1$, and 10,000 shots per circuit.

Figure~\ref{fig:gpu_scaling} summarizes both the observable-estimation consistency and execution-time scaling across the evaluated qubit widths. The top row compares the sampled expectation values obtained using 1, 4, 8, and 16 GPUs. Across all three Hamiltonians, the results from the different GPU configurations remain closely aligned, indicating that distributing independent measurement circuits across multiple GPUs does not introduce a systematic change in the sampled expectation values. This consistency demonstrates that the parallel execution strategy preserves the observable-estimation behavior across GPU configurations while enabling scalable execution.

The bottom row of Fig.~\ref{fig:gpu_scaling} reports execution time as a function of qubit width on a logarithmic scale. Increasing the GPU count consistently reduces execution time, although the achievable improvement depends on the number of independent commuting groups generated for each Hamiltonian. TFIM contains only two commuting groups and therefore has a theoretical circuit-level speedup limit of $2\times$. In contrast, the Bose-Hubbard Hamiltonian contains 9--11 groups, enabling substantially greater parallelism, while H2 contains between 54 and 1,251 groups over the evaluated qubit widths and provides sufficient independent work to utilize all 16 GPUs effectively.

Table~\ref{tab:gpu-observable-results} summarizes the results at the largest qubit width evaluated for each Hamiltonian. TFIM achieves a speedup of $1.9\times$, approaching its two-group theoretical limit. Bose-Hubbard achieves an $8.2\times$ speedup with nine commuting groups at 28 qubits, corresponding to approximately 91\% of the achievable maximum of $\min(16,9)=9$. H2 exhibits the greatest improvement, reducing execution time from 283.81~s on one GPU to 20.51~s on 16 GPUs, corresponding to a $13.8\times$ speedup and approximately 86\% efficiency relative to the ideal $16\times$ speedup.

The scaling behavior observed across all three Hamiltonians is governed primarily by the relationship between the number of commuting measurement groups, $C$, and the number of available GPUs, $G$. When $C \gg G$, the measurement circuits can be distributed across all GPUs, allowing near-linear scaling. Conversely, when $C < G$, some GPUs remain idle and the achievable speedup is fundamentally bounded by $C$. As the problem size increases, the cost of simulating each measurement circuit dominates fixed overheads such as circuit construction, MPI communication, result aggregation, and classical post-processing, making circuit-level parallelism increasingly effective. Consequently, the commuting-group structure of the Hamiltonian determines both the available parallelism and the attainable performance on multi-GPU systems.


\subsubsection{Hybrid Multi-GPU Parallelism}

To demonstrate the value of hybrid parallelism, we simulated with a 36-qubit Bose-Hubbard Hamiltonian that requires a minimum of 16 GPUs to fit in memory on Perlmutter. With only 16 GPUs, the execution time is 9.8 hours. By running on 192 GPUs with the \texttt{--gpus\_per\_circuit=16} option, this time is reduced by 11.9X to 50 minutes, realizing 99\% of the ideal speedup. This is significantly higher efficiency than would be expected from \texttt{mgpu} parallelism alone, due to the parallel evaluation of commuting groups with consistent individual simulation times. 

For the results evaluated here at smaller qubit counts, this distributed statevector mode (\texttt{mgpu}), in which all GPUs cooperatively simulate a single circuit, provided more modest speedups of $1.4\times$--$2.9\times$ on the same workloads. As the parallel efficiency decreases with circuit width, CUDA-Q will default to running on a single GPU when using less than 26 qubits, even if more GPUs are provided for the run. Indeed, our results in which this default behavior is bypassed, discussed in Appendix~\ref{app:mgpu_scaling}, result in slower parallel execution times. Therefore, this additional mode of hybrid parallelism can be a significant efficiency benefit when simulating many independent circuits or multiple QPUs.

\subsection{Evaluation on Quantum Hardware}

We next evaluate the partition-based execution strategy on IBM quantum hardware (\texttt{ibm\_boston}). Unlike GPU simulation, these experiments assess the reduction in quantum execution time together with the fidelity implications of executing multiple circuits simultaneously on disjoint qubit partitions.

\subsubsection{Standard QED-C Benchmark Algorithms}

We evaluated the partition-based execution strategy on IBM Boston using Bernstein--Vazirani (BV), the transverse-field Ising model (TFIM), and the Quantum Fourier Transform (QFT). For each algorithm, we compare conventional sequential execution with parallel execution over a range of circuit widths. Figure~\ref{fig:fidelity_and_time_boston} summarizes the resulting Hellinger fidelity, transpiled circuit depth, billed execution time, and parallel speedup.

The top row of Fig.~\ref{fig:fidelity_and_time_boston} compares the Hellinger fidelity for sequential and parallel execution together with the corresponding transpiled circuit depth. Across all three benchmarks, parallel execution generally produces lower fidelity because partitioning restricts the transpiler to a subset of physical qubits, which can increase routing and the number of two-qubit gates. The impact of this depends strongly on circuit depth. BV maintains relatively high fidelity because its transpiled depth remains nearly constant. TFIM shows increasing depth and a corresponding decline in fidelity, with the parallel implementation degrading more rapidly at larger widths. QFT shows the strongest effect: its depth increases rapidly with circuit width, reaching nearly 280 layers at 16 qubits, and the fidelity of both sequential and parallel execution approaches zero.

The bottom row reports the billed execution time and the reduction achieved through parallel execution. Across all three benchmarks, executing independent circuits simultaneously on disjoint qubit partitions consistently reduces billed QPU time. BV achieves approximately 70--75\% reduction over most circuit widths, with the benefit decreasing for the largest circuits as fewer partitions remain available. TFIM maintains approximately 65--75\% reduction across the evaluated range. QFT shows a comparable reduction for smaller circuits, but the benefit decreases at larger widths as the number of available partitions is reduced.

Overall, these results show the tradeoff between execution efficiency and solution quality on current hardware. Partition-based execution substantially reduces billed QPU time, but constraining circuits to disjoint qubit regions can reduce fidelity, particularly for deeper circuits with greater routing requirements. The approach is therefore most effective for workloads containing many shallow or moderately deep circuits, where significant execution-time reductions can be achieved without a large loss in fidelity.


\begin{figure*}[]
    \includegraphics[width=1\textwidth]{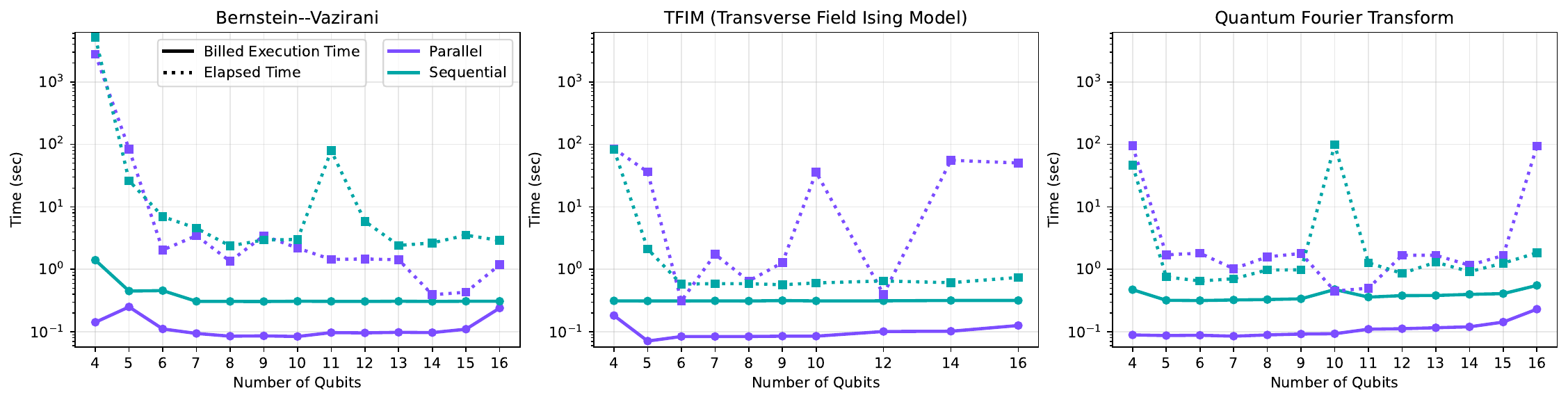}
    \par\medskip
    \includegraphics[width=1\textwidth]{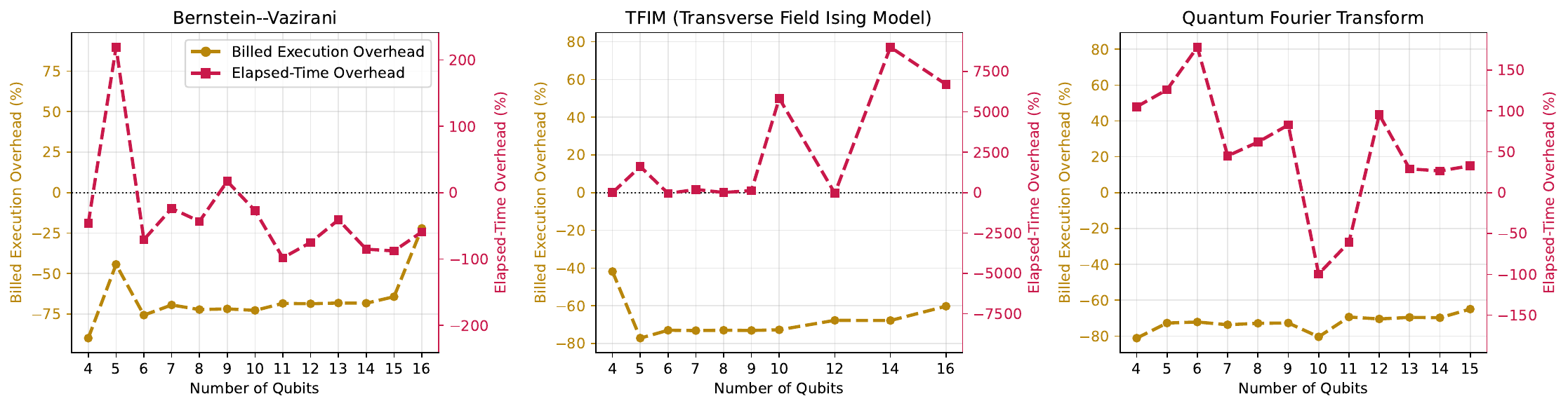}
    \caption{\textbf{Execution-Time and Overhead Analysis for Standard QED-C Benchmarks on IBM Boston.} The top row compares billed QPU execution time (solid lines) and total elapsed time (dotted lines) for sequential and parallel execution of BV, TFIM, and QFT. The bottom row reports the corresponding percentage change relative to sequential execution. Parallel execution consistently reduces billed QPU time, while elapsed time is more variable because it also includes compilation, orchestration, and service-level overhead.}

    \label{fig:time_comparison_boston}
\end{figure*}

\subsubsection{Observable Estimation}

Figure~\ref{fig:hamlib_expectation_and_time_boston} compares observable-estimation accuracy and billed expectation-computation time for sequential and parallel execution of the TFIM, Bose-Hubbard, and H2 Hamiltonians. All experiments use simple sampling with $K=1$ Trotter step, evolution time $t=0.1$, and 10,000 shots per measurement circuit. Billed time is obtained from IBM's \texttt{execution\_spans} metadata and excludes queue wait and transpilation overhead.

Parallel execution reduces billed QPU time for all three Hamiltonians, with the achievable reduction determined primarily by the number of independent measurement circuits and the number of disjoint partitions available on the processor. TFIM contains only two commuting measurement groups and therefore provides at most two-way circuit parallelism. Bose-Hubbard produces 9--12 measurement circuits over the evaluated range and more clearly demonstrates the effect of partition availability. At 4 qubits, nine partitions allow all nine circuits to execute in a single round. At 8 qubits, six partitions require two rounds for eleven circuits, while at 10 qubits five partitions require three rounds. This increase in execution rounds accounts for the decreasing billed-time reduction as circuit width increases. H2 contains substantially more measurement groups and therefore provides ample circuit-level parallelism, although the number of available partitions again decreases as circuit width increases.

The expectation values obtained through parallel execution generally remain close to those obtained through sequential execution. In several cases, the parallel result is closer to the exact value, consistent with the calibration-aware partition mapper selecting lower-error regions of the processor. This improvement should not be attributed to parallel execution itself. For larger circuits, the smaller number of feasible partitions and the additional routing required within those partitions can instead increase the deviation from the exact result.

These results show the primary tradeoff for partition-based observable estimation on current hardware. Parallel execution can substantially reduce billed QPU time by making use of otherwise idle qubit regions, but the available parallelism decreases as circuit width increases. Performance and accuracy therefore depend on both the number of available partitions and the quality and connectivity of the qubit regions used for execution.

\vspace{0.3cm}
\subsubsection{Timing Analysis}

To distinguish reductions in QPU usage from changes in end-to-end runtime, we separately analyze billed QPU execution time and total elapsed wall-clock time. Billed time is obtained from IBM's \texttt{execution\_spans} metadata and represents the QPU execution interval, while elapsed time also includes circuit preparation, transpilation, network communication, job submission, scheduling, and result retrieval. For the parallel path, additional overhead arises from composing and transpiling multiple sub-experiments into composite circuits spanning disjoint regions of the processor.

Figure~\ref{fig:time_comparison_boston} compares these two timing measures, with overhead reported as the percentage change relative to sequential execution.
The top row reports the raw timing values, with solid lines denoting billed execution time and dotted lines denoting elapsed time. The bottom row presents the corresponding parallel overhead relative to sequential execution. We define the overhead for a timing quantity $T$ as
\begin{equation}
\mathrm{Overhead}(T)
=
\frac{T_{\mathrm{parallel}}-T_{\mathrm{sequential}}}
     {T_{\mathrm{sequential}}}
\times 100\%.
\end{equation}
Negative overhead therefore indicates that parallel execution reduces the corresponding time, whereas positive overhead indicates an increase. Table~\ref{tab:qpu_speedup_summary} summarizes these results by reporting the average billed execution speedup and elapsed-time factor across all evaluated qubit sizes for each benchmark.

\begin{table}
\centering
\caption{Average billed execution speedup and elapsed-time factor for the parallel execution framework on IBM quantum hardware. Values are averaged across all evaluated qubit sizes for each benchmark. An elapsed-time factor below 1 indicates a reduction in wall-clock time.}
\label{tab:qpu_speedup_summary}

\begin{tabular}{@{}lcc@{}}
\toprule
\textbf{Benchmark} & \textbf{Billed Execution Speedup} & \textbf{Elapsed-Time Factor} \\
\midrule
BV & 3.58$\times$ & 0.68$\times$ \\
TFIM & 3.33$\times$ & 24.42$\times$ \\
QFT & 3.64$\times$ & 1.52$\times$ \\
\bottomrule
\end{tabular}

\end{table}

Across all three benchmarks, parallel execution consistently reduces billed QPU execution time. The billed-time reduction is approximately $60$--$80\%$ for most TFIM and QFT instances, with BV showing a similar reduction over most circuit widths. As summarized in Table~\ref{tab:qpu_speedup_summary}, the average billed execution speedups are $3.58\times$, $3.33\times$, and $3.64\times$ for BV, TFIM, and QFT, respectively. These results confirm that executing independent circuits concurrently on disjoint partitions substantially reduces billable QPU usage.

Elapsed wall-clock time shows considerably greater variability. BV generally benefits from parallel execution, with an average elapsed-time factor of $0.68\times$, while QFT averages $1.52\times$. TFIM exhibits much larger elapsed time, averaging $24.42\times$ the sequential execution time, with large spikes at selected circuit widths. These increases do not correspond to comparable increases in billed QPU time and instead arise primarily from classical compilation and orchestration. Partition selection itself requires only 1--2 seconds, indicating that the much larger elapsed-time overhead arises primarily from subsequent composite-circuit construction and transpilation rather than from the partition-finding algorithm. In particular, \texttt{ParallelExperiment} must compose the mapped sub-experiments and transpile the resulting physical circuit while preserving the assigned partition boundaries, and this processing can dominate the relatively short QPU execution interval.

The timing results therefore show a clear distinction between QPU-resource efficiency and end-to-end runtime. Partition-based execution consistently reduces billed QPU time, but these savings do not necessarily translate into lower wall-clock latency with the current implementation. Much of this overhead is associated with the surrounding compiler and runtime path rather than the parallel QPU execution itself, suggesting that improvements in composite-circuit construction, transpilation, and reuse of previously determined partition mappings could allow the reduction in QPU execution time to translate more directly into reduced end-to-end runtime.

\section{Discussion}
\label{sec:discussion}

\subsection{Workload Structure Determines Parallel Speedup}
\label{subsec:discussion_structure}

Across both experiment classes, speedup is determined primarily by the number and cost of independent circuits available for concurrent execution.
For observable estimation, this is determined by the number of commuting Pauli groups. TFIM, despite having 40--56 terms at the qubit widths tested, groups into only 2 measurement circuits because of the regular structure of its nearest-neighbor interactions. H2, with 185--2,951 terms, produces 54--1,251 groups due to the more complex interactions in the molecular Hamiltonian. Hamiltonians with many commuting groups can therefore benefit substantially from multi-GPU or multi-QPU parallelism, while models producing only a few groups have limited circuit-level parallelism regardless of the available resources.

An important practical consequence is that the choice of grouping algorithm affects not only measurement cost and statistical variance, but also the parallelism available during execution. More aggressive grouping strategies that minimize circuit count can simultaneously reduce the opportunity for parallel execution. A complete optimization should therefore balance grouping efficiency against the parallel resources available for execution.

\subsection{Scaling Efficiency}
\label{subsec:discussion_scaling}

Scaling across the standard QED-C and observable-estimation experiments depends on both circuit cost and the amount of available circuit-level parallelism.
For the Bose-Hubbard Hamiltonian on GPU simulators, parallel efficiency improves with qubit width, reaching $8.2\times$ at 28 qubits with a theoretical maximum of $\min(16,9)=9\times$. This trend arises because MPI coordination cost is approximately independent of qubit width, while the cost of executing each circuit grows rapidly. As circuit width increases, execution dominates the coordination overhead, allowing the available circuit-level parallelism to be used more efficiently.

H2 illustrates the complementary effect of having many independent measurement circuits. At 20 qubits, its 1,251 measurement circuits provide sufficient parallel work to achieve a $13.8\times$ speedup on 16 GPUs, or 86\% of the theoretical maximum. These results show that efficient scaling requires having enough work within each circuit and enough independent circuits to keep the available GPUs fully occupied.

For circuits that exceed the memory capacity of a single GPU, circuit-level parallelism can be combined with distributed statevector simulation. In the 36-qubit Bose-Hubbard experiment, each circuit was distributed across 16 GPUs while independent measurement groups were executed concurrently across 192 GPUs, reducing execution time from approximately 9.8 hours to 50 minutes. This hybrid approach extends circuit-level parallelism to workloads for which individual circuits themselves require multiple GPUs.

On quantum hardware, parallelism is currently limited by the number of sufficiently large, low-error partitions available on the processor. On the 156-qubit device used here, this generally limited execution to approximately 4--6 simultaneous circuits, resulting in lower speedups than those observed on GPUs. Larger devices with improved gate fidelity should support more usable partitions, allowing hardware execution to benefit from greater circuit-level parallelism.

\subsection{Fidelity and Utilization Tradeoff on Hardware}
\label{subsec:discussion_fidelity}

Parallel execution on hardware introduces a tradeoff between device utilization and circuit fidelity. Our error-scored partition mapping mitigates this by selecting the highest-quality regions first, but additional partitions generally require the use of progressively lower-quality qubit regions. On \texttt{ibm\_boston}, for example, the sixth 8-qubit partition has an average gate error approximately $15\times$ higher than the first. Restricting circuits to individual partitions can also increase routing requirements. This is especially true when the available connectivity within a partition is lower than that available to sequential execution.

These effects are visible in the observed fidelity retention of 83--92\%. The lower end corresponds to deeper circuits, where routing constraints and accumulated two-qubit gate errors have a greater impact. Shallower circuits generally retain a larger fraction of the sequential fidelity. For very deep circuits such as the larger QFT instances, the fidelity of both sequential and parallel execution approaches zero. This indicates that circuit depth and hardware noise become the dominant limitation, rather than parallel execution.

The gap parameter provides additional control over the tradeoff between partition count and isolation. A gap of 2, used by default in these experiments, reduces the number of available partitions while providing greater separation between simultaneously executing circuits. Reducing the gap to 1 increases the number of available partitions by approximately 30\%, but also reduces this separation. We did not observe significant crosstalk effects with a gap of 2 on \texttt{ibm\_boston}. However, a systematic study of crosstalk as a function of partition separation, circuit depth, and simultaneous partition count remains an important area for future work.

\subsection{Forward Applicability}
\label{subsec:discussion_forward}

While our results focus primarily on standardized benchmarks and observable estimation for Hamiltonian simulation, the parallel execution approach applies more generally to workloads producing many similarly sized independent circuits. Examples include variational algorithms such as VQE and QAOA, quantum machine learning ensembles, randomized benchmarking and tomography protocols, and parameter-sweep workflows at fixed circuit width.

The partition mapping algorithm requires a coupling map and calibration information for the underlying hardware. Although we have validated the approach only on IBM superconducting hardware, the method is not inherently IBM-specific and could be adapted to other architectures that provide corresponding topology and error information. Different hardware technologies may require different partitioning strategies. For example, trapped-ion systems present a different connectivity structure from the sparse coupling graphs considered here. Cross-platform validation remains an area for future work.

The potential benefit of parallel execution should increase as quantum processors become larger and more reliable. Consider, for example, a plausible future scenario in which processors contain 500--1,000 physical qubits and improvements in gate fidelity allow circuits of 40--50 qubits to execute reliably. Such a device could potentially support 10--20 disjoint regions for concurrent circuit execution, depending on circuit width, device topology, and the quality of the available qubits. Both developments work in favor of parallel execution: additional physical qubits provide more potential partitions, while improved fidelity increases the size and complexity of circuits that can execute reliably within each partition. The GPU results demonstrate the level of circuit-parallel speedup that becomes possible when this degree of parallelism is available, providing an indication of the performance that larger and more reliable quantum processors could eventually support.

\subsection{Future Optimizations}
\label{subsec:discussion_optimizations}

Several improvements could increase the efficiency and fidelity of hardware parallel execution. Partition selection could account more directly for routing requirements and measured crosstalk. Allocating additional qubits around each partition could provide the transpiler with greater routing flexibility. Dynamically adjusting the separation between partitions based on measured crosstalk could allow more effective use of the available hardware.

There are also opportunities to reduce the setup and compilation overhead associated with parallel execution. Partition layouts for specific backends and circuit widths could be precomputed and reused, eliminating the 1--2 s partition-selection cost. For workloads executed repeatedly with different parameters, portions of the compilation process may also be reusable. For example, Hamiltonian simulation circuits could potentially be compiled once, with measurement rotations or other run-time parameters varied between executions. Such optimizations could help translate the reductions in billed QPU execution time demonstrated here into corresponding reductions in end-to-end wall-clock time.

\section{Summary and Conclusions}
\label{sec:conclusion}

Current quantum processors often contain far more physical qubits than can be reliably used together for general application circuits. We have demonstrated that this excess capacity can be productively exploited today through parallel circuit execution, reducing execution time and billed QPU usage without requiring any hardware modification.

Across standard QED-C and observable-estimation benchmarks, GPU results show that circuit-level speedup depends on the number and cost of independent circuits. H2, with 1,251 independent measurement circuits, achieved a $13.8\times$ speedup on 16 GPUs, representing 86\% of the theoretical maximum. Bose-Hubbard achieved an $8.2\times$ speedup with nine measurement groups, while TFIM, whose simple structure yields only two groups, was limited to $1.9\times$. These results show that the structure of the workload, and particularly the number of independent circuits it produces, determines how much parallelism can be exploited. For circuits too large for a single GPU, circuit-level parallelism can also be combined with distributed statevector simulation; for the 36-qubit Bose-Hubbard calculation, this reduced execution time from approximately 9.8 hours to 50 minutes using 192 GPUs.

On the 156-qubit IBM \texttt{ibm\_boston} processor, our error-aware partition mapping identifies disjoint qubit regions that allow multiple independent circuits to execute simultaneously. Across the workloads evaluated, parallel execution produced approximately $2$--$3\times$ reductions in billed QPU execution time while retaining 83--92\% of the sequential-execution fidelity for circuits that could be executed reliably on the hardware. The results also expose the principal cost of this approach: constraining circuits to disjoint regions can reduce fidelity through less favorable qubit selection and routing, while the current software path can introduce substantial compilation and orchestration overhead.

The GPU and hardware experiments reveal two sides of the same opportunity. On current quantum hardware, the number and quality of usable partitions limit the parallelism that can be achieved. The GPU results show the performance ceiling when those constraints are removed: with sufficient independent circuits and computational resources, circuit-level execution can approach the available parallel capacity. As quantum processors grow toward 500--1,000 physical qubits and improvements in fidelity make reliable 40--50 qubit circuits possible, the same approach could potentially support 10--20 simultaneous circuit partitions. More physical qubits provide more opportunities for parallel execution, while better fidelity makes each partition more useful.

Both execution modes have been integrated into the QED-C Application-Oriented Benchmark suite, allowing the same application workloads to execute sequentially or in parallel without changing their circuit-generation or analysis logic. The approach is not limited to any specific application: standard benchmarks, observable estimation, variational algorithms, parameter sweeps, and other workloads generating independent circuits can exploit the same parallelism. Parallel circuit execution therefore provides a way to extract more useful computation from the resources already available, while providing a path to substantially greater parallelism as quantum processors become larger and more capable.

\section*{Code Availability}

The quantum application programs described in this work are implemented in the QED-C open-source Application-Oriented Benchmark suite, available at \url{https://github.com/SRI-International/QC-App-Oriented-Benchmarks}.
Execution scripts, data collection tools, and visualization programs used to generate the results reported here are available at \url{https://github.com/quantumcomputingdata/QC-App-Benchmarks-Data}.

\section*{Acknowledgements}

We thank David E. Bernal Neira, Daan Camps, Sonika Johri, Efekan Kokcu, Debarshi Kundu, Neer Patel, and Anish Giri for valuable review and commentary that contributed to this manuscript. Many other members of the QED-C member organization and the Committee on Standard and Performance Metrics also provided valuable input into the direction of this work.

AC and WAdJ were supported in part by the U.S. Department of Energy, Office of Science, National Quantum Information Science Research Centers, Quantum Systems Accelerator (Award No. DE-SCL0000121), under Contract No. DE-AC02-05CH11231. AC was additionally supported in part by the National Energy Research Scientific Computing Center (NERSC), a U.S. Department of Energy Office of Science User Facility (Project No. m888). SN was supported by the University of Central Florida.

This research used resources of the National Energy Research Scientific Computing Center, a DOE Office of Science User Facility supported by the Office of Science of the U.S. Department of Energy under Contract No. DE-AC02-05CH11231  using NERSC awards NERSC DDR-ERCAP0038362, NERSC DDR-ERCAP0034389 and NERSC DDR-ERCAP0038885.

We acknowledge the use of IBM Quantum services for this work under NERSC DDR-ERCAP0038362 and NERSC DDR-ERCAP0038885. The views expressed in this manuscript are those of the authors and do not reflect the official policy or position of IBM or the IBM Quantum team.
IBM Quantum. https://quantum-computing.ibm.com, 2026.



\bibliographystyle{IEEEtran}
\bibliography{references}

\clearpage
\appendices

\section{Topology-Aware Partition Mapping: Algorithm Details}
\label{app:partition_mapping}

This appendix provides detailed algorithmic descriptions of the partition finding and circuit assignment procedures, intended for researchers who may wish to reproduce or extend this work.

\subsection{Subgraph Growth}
\label{app:subgraph_growth}

The partition-finding algorithm begins by constructing an undirected graph $G = (V, E)$ from the backend's coupling map, where vertices represent physical qubits and edges represent available two-qubit gates.
For each starting vertex $v_0 \in V$, we grow a connected subgraph of target width $w$ using the following procedure:

\begin{algorithm}
\caption{Greedy Subgraph Growth}
\label{alg:subgraph_growth}
\begin{algorithmic}[1]
\Require Graph $G = (V, E)$, starting vertex $v_0$, target width $w$
\Ensure Connected subgraph $S \subseteq V$ with $|S| = w$
\State $S \gets \{v_0\}$
\While{$|S| < w$}
    \State $F \gets \{u \in V \setminus S : \exists s \in S, (u,s) \in E\}$ \Comment{Frontier}
    \If{$F = \emptyset$}
        \State \textbf{fail} \Comment{Cannot grow to target width}
    \EndIf
    \State $u^* \gets \arg\max_{u \in F} |\{s \in S : (u,s) \in E\}|$ \Comment{Most connections to cluster}
    \State $S \gets S \cup \{u^*\}$
\EndWhile
\State \Return $S$
\end{algorithmic}
\end{algorithm}

The growth heuristic (line 7) selects the frontier node with the most connections to the existing cluster.
This produces compact, well-connected subgraphs rather than chains, which is important for two reasons: compact subgraphs have smaller diameter (reducing SWAP overhead during transpilation), and more internal edges provide the transpiler with greater routing flexibility.

On heavy-hex topologies (used by IBM Eagle and Heron processors), where most qubits have degree 2--3, this heuristic tends to produce T-shaped or cross-shaped subgraphs rather than linear chains.

\subsection{Multi-Criteria Scoring}
\label{app:scoring}

Each candidate subgraph $S$ is scored by a tuple $(e_S, d_S, -|E_S|)$ where:

\begin{itemize}
    \item $e_S = \frac{1}{|E_S|} \sum_{(u,v) \in E_S} \epsilon_{uv}$ is the average two-qubit gate error on edges within $S$, where $\epsilon_{uv}$ is read from the backend calibration data for the native two-qubit gate operations.
    \item $d_S = \max_{u,v \in S} \text{dist}_S(u,v)$ is the diameter of the subgraph---the maximum shortest path length between any pair of vertices within $S$.
    \item $|E_S| = |\{(u,v) \in E : u \in S, v \in S\}|$ is the internal edge count, negated so that more edges sorts earlier.
\end{itemize}

Candidates are sorted in ascending order by this tuple. The error score dominates: on \texttt{ibm\_boston}, average two-qubit gate errors range from 0.002 (best 8-qubit partition) to 0.149 (worst), a $70\times$ spread that dwarfs differences in diameter or edge count.

\subsection{Greedy Non-Overlapping Selection with Gap}
\label{app:selection}

From the sorted candidate list, partitions are selected greedily:

\begin{algorithm}
\caption{Greedy Partition Selection}
\label{alg:partition_selection}
\begin{algorithmic}[1]
\Require Sorted candidates $C = [(S_1, \text{score}_1), \ldots]$, target count $P$, gap $g$, max error $\epsilon_{\max}$
\Ensure Disjoint partitions $\mathcal{P} = \{S_{i_1}, S_{i_2}, \ldots\}$
\State $\mathcal{P} \gets \emptyset$, $\text{used} \gets \emptyset$
\For{$(S, \text{score})$ in $C$}
    \If{$|\mathcal{P}| \geq P$}
        \textbf{break}
    \EndIf
    \If{$e_S > \epsilon_{\max}$}
        \State Log warning, \textbf{continue} \Comment{Reject high-error partition}
    \EndIf
    \State $N_g(S) \gets \{v \in V : \min_{s \in S} \text{dist}_G(v,s) \leq g\}$ \Comment{Gap neighborhood}
    \If{$N_g(S) \cap \text{used} = \emptyset$}
        \State $\mathcal{P} \gets \mathcal{P} \cup \{S\}$
        \State $\text{used} \gets \text{used} \cup N_g(S)$
    \EndIf
\EndFor
\If{$|\mathcal{P}| < P$ and $g > 0$}
    \State Retry with $g \gets g - 1$ \Comment{Gap retry}
\EndIf
\State \Return $\mathcal{P}$
\end{algorithmic}
\end{algorithm}

The gap parameter $g$ controls the separation between partitions.
With $g=2$ (default), any qubit within 2 hops of a selected partition is marked as used, preventing adjacent partitions from sharing nearby qubits.
The gap retry strategy (line 14) ensures that we find a reasonable number of partitions even on devices where gap=2 is too restrictive.

\subsection{Multi-Width Partition Allocation}
\label{app:multi_width}

For benchmarks that produce circuits of varying widths, we allocate partitions in two phases:

\textbf{Phase 1: One partition per width.}
Sort unique widths in decreasing order.
For each width $w_i$, find one partition of size $w_i$ on the device (using Algorithm~\ref{alg:partition_selection} with $P=1$), excluding qubits already allocated.
Stop at the first width that cannot be accommodated---this prevents allocating tiny partitions that waste device space while leaving large-width groups bottlenecked.

\textbf{Phase 2: Additional partitions.}
For each width that received a partition in Phase~1, attempt to find additional partitions on the remaining device qubits.
This reduces the number of rounds: if a width has $n$ circuits and $p$ partitions, it requires $\lceil n/p \rceil$ rounds.

After partition allocation, circuits are assigned in two passes:
\begin{enumerate}
    \item \textbf{Exact match:} Circuits whose width matches a partition's width are assigned directly, round-robin across that width's partitions.
    \item \textbf{Padded distribution:} Remaining circuits are padded with idle qubits to match partition widths, then distributed round-robin across \emph{all} partitions. Distributing across all partitions (rather than only the smallest) prevents overloading a single partition when many circuits are smaller than the available partition sizes.
\end{enumerate}

\subsection{Transpiler Escape and Pre-Transpile Fallback}
\label{app:pretranspile}

The \texttt{ParallelExperiment} framework enforces that each sub-experiment's transpiled circuit uses only the assigned \texttt{physical\_qubits}.
When the Qiskit transpiler, given full device connectivity, routes a circuit through qubits outside the assigned partition, a \texttt{QiskitError} is raised.

Our fallback procedure handles this by constructing a restricted coupling map for each partition:

\begin{enumerate}
    \item Extract all edges $(u,v)$ from the full coupling map where both $u, v \in S_i$ (the partition's qubits).
    \item Remap to local indices: if $S_i = \{q_3, q_7, q_{12}\}$, map to $\{0, 1, 2\}$.
    \item Construct a \texttt{CouplingMap} from the remapped edges.
    \item Transpile each circuit onto this restricted map using \texttt{optimization\_level=1}.
    \item Run \texttt{ParallelExperiment} with the pre-transpiled circuits and \texttt{optimization\_level=0} (skip re-transpilation).
\end{enumerate}

This guarantees partition containment at the cost of constraining the transpiler's search space.
On \texttt{ibm\_boston}, the fidelity impact varied across workloads, including a reduction from 0.920 to 0.77 for a 4-qubit QFT instance where pre-transpilation was triggered unnecessarily during early testing, while an 8-qubit HamLib instance retained approximately 83\% of the sequential fidelity.

The escape problem is inherent to using a full-device transpiler with partition constraints.
Future hardware compilers that accept explicit routing boundaries could eliminate this issue entirely.

\subsection{Implementation Notes}

\begin{itemize}
    \item \textbf{NetworkX dependency:} The partition-finding algorithm uses NetworkX for graph operations (subgraph construction, diameter computation, shortest paths). This adds a lightweight dependency but simplifies the implementation significantly.
    \item \textbf{Qiskit Experiments dependency:} \texttt{ParallelExperiment} and \texttt{BaseExperiment} from the \texttt{qiskit-experiments} package provide the compositing and result decomposition infrastructure~\cite{qiskit_experiments}.
    \item \textbf{Result reordering:} An \texttt{assignment\_map} tracks the mapping from $(partition, index)$ to original circuit order, enabling correct result reconstruction after \texttt{ParallelExperiment}'s decomposition.
    \item \textbf{Timing extraction:} On IBM hardware, per-circuit execution time is extracted from the \texttt{execution\_spans} metadata in the \texttt{PrimitiveResult}, divided by the number of circuits in the job.
\end{itemize}


\section{GPU-Based Parallel Circuit Execution Details}
\label{app:multi_gpu_details}

Direct GPU-based statevector simulation can generate exact expectation values more efficiently than sampling. For our hardware experiments, however, circuit execution uses finite-shot sampling, consistent with the execution model described in Section~\ref{subsec:multigpu}. This provides a common basis for comparing parallel execution in the two environments.

For GPU-based execution, $T$ MPI ranks are divided into $G=T/P$ independent subcommunicators, where $P$ is the number of GPUs assigned to each circuit. Measurement circuits are distributed in approximately equal contiguous blocks among the $G$ groups. Each group executes its assigned circuits independently, with no communication between groups during circuit execution. The group leaders then return their results to the world leader, which restores the original circuit order and performs the standard post-processing and analysis.

When $P=1$, each circuit executes on a single GPU, providing circuit-level parallelism across the available GPUs. When $P>1$, CUDA-Q distributed statevector simulation is used within each group while multiple groups execute different circuits concurrently, providing hybrid parallelism across and within circuits. The value of $P$ is controlled by the \texttt{--gpus\_per\_circuit} option.

While CUDA-Q has supported the \texttt{mqpu} option for execution of multiple circuits in parallel, for the QED-C benchmark suite, we have implemented support using the new MPI subcommunicator functionality. This allows fine-grained control of the MPI implementation for parallelism with finite-shot sampling and allows for a fully portable solution for hybrid parallelism with multiple circuits executed in parallel, each using multiple GPUs for simulation.

The parallel circuit execution mode is activated by passing the \texttt{--parallel} flag to the \texttt{qedclib} program invocation. Given $T$ MPI tasks and $P$ GPUs per circuit, the procedure forms $G=T/P$ independent subcommunicators and distributes the measurement circuits among them. 

When this mode is active, the following execution flow is used:

\begin{enumerate}[label=\alph*.]
\item Given $P$ as set by \texttt{-{-}gpus\_per\_circuit} and the total number of MPI tasks, $T$, split \texttt{MPI\_COMM\_WORLD} into $G = T/P$ subcommunicators, using the CUDA-Q \texttt{set\_communicator()} function such that each circuit is parallelized across $P$ ranks/GPUs within each subcommunicator. Each subcommunicator assigns an MPI rank as a \emph{circuit leader} with a single \emph{world leader} as rank 0.

\item Each rank determines the unique subset of $C$ measurement circuits it will handle based on its subcommunicator ID.

\item Each rank receives a contiguous block of approximately $\lceil C / G \rceil$ circuits.

\item Each subcommunicator executes its assigned circuits independently and locally, with no communication between subcommunicators during circuit execution.

\item After all circuits complete, each circuit leader returns its partial results to the world leader via an MPI gather operation (\texttt{mpi.gather()}).

\item The world leader reassembles the full result set in the original circuit order and performs the standard post-processing to compute expectation values.
\end{enumerate}

Only the world leader executes subsequent analysis steps, ensuring that result processing occurs exactly once. The contiguous block distribution strategy is simple and effective when circuits have roughly uniform execution times; the implications of non-uniform execution times are discussed in Section~\ref{sec:discussion}.

The key implementation changes to the \texttt{qedclib} framework were confined to three files: the CUDA-Q execution module (\texttt{qedclib/cudaq/execute.py}), which received the new \texttt{\_execute\_parallel\_mpi()} function and associated block index utilities; the MPI wrapper module (\texttt{qedclib/qcb\_mpi.py}), which received the new subcommunicator functionality and the \texttt{gather()} MPI communication function; and the HamLib benchmark entry point (\texttt{hamlib/hamlib\_simulation\_benchmark.py}), which received the \texttt{-{-}pm} and \texttt{-{-}gpus\_per\_circuit} command-line arguments and the leader check logic following parallel execution. The Qiskit execution module was also updated for API compatibility, positioning the framework for future execution on quantum hardware via the Qiskit backend.

\section{Detailed GPU Simulation Results}
\label{app:gpu_detailed}

\begin{figure*}[t!]
  \centering
  \includegraphics[width=1.95\columnwidth]{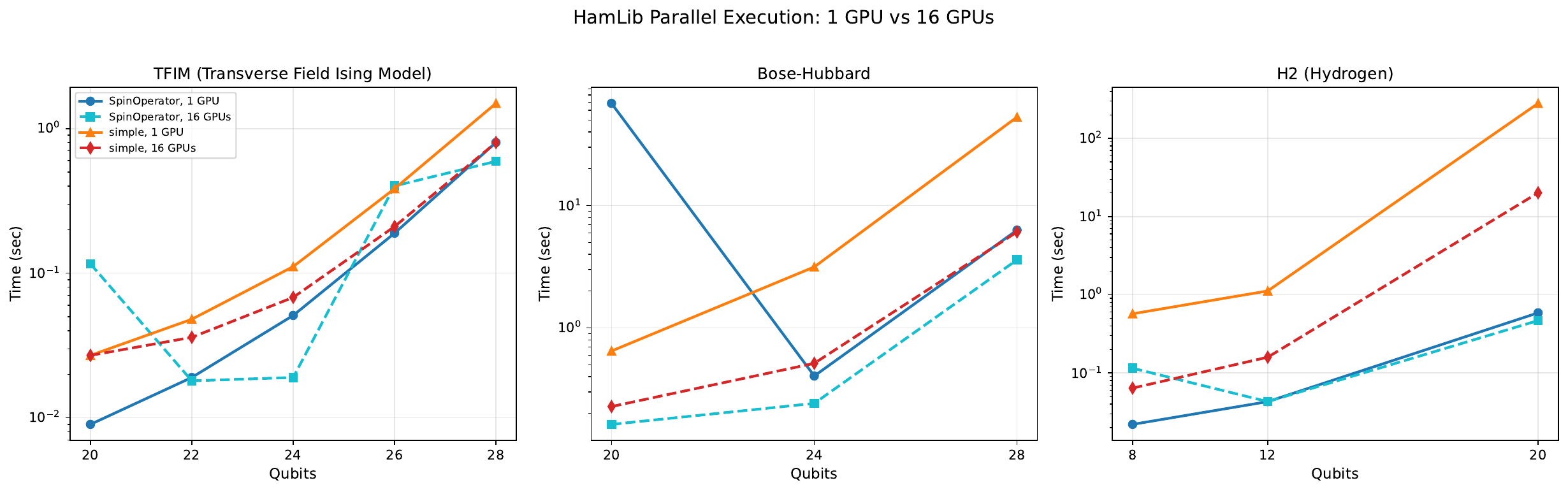}
  \caption{\textbf{Comparison of CUDA-Q execution modes for Hamiltonian observable estimation.}
  Each panel displays execution time (log scale) versus qubit width across three benchmark Hamiltonians (TFIM, Bose-Hubbard, H2). Four configurations are shown: SpinOperator on a single GPU (baseline), SpinOperator with \texttt{mgpu} statevector distribution across 16 GPUs, simple Pauli sampling on a single GPU, and simple sampling with \texttt{-{-}parallel} parallel circuit execution across 16 GPUs. The TFIM Hamiltonian, which produces only 2 measurement circuits, achieves near-maximum theoretical speedup of $\sim$1.9$\times$. The H2 chemistry Hamiltonian, with 1,251 circuits at 20 qubits, achieves 13.8$\times$ speedup on 16 GPUs, approaching the theoretical maximum of 16$\times$.}
  \label{fig:parallel_exec}
\end{figure*}

This appendix provides per-Hamiltonian results across all qubit widths tested, including comparison between the SpinOperator (exact) and simple sampling (shot-based) computation methods.

\subsection{Observable Computation Methods}
\label{subsec:obs_methods}

\begin{table}
\centering
\caption{Execution Configurations for HamLib Observable Benchmarks}
\label{tab:configs}
\begin{tabular}{@{}lll@{}}
\toprule
\textbf{Configuration} & \textbf{Grouping} & \textbf{Multi-GPU Mode} \\
\midrule
SpinOperator, 1 GPU     & \texttt{SpinOperator} & None (baseline) \\
SpinOperator, mgpu      & \texttt{SpinOperator} & mgpu (statevector dist.) \\
Simple, 1 GPU           & \texttt{simple}       & None (baseline) \\
Simple, -pm            & \texttt{simple}       & Parallel circuit exec. \\
Hybrid, -gpc \emph{N}  & \texttt{simple}       & Parallel SV and circuit exec. \\
\bottomrule
\end{tabular}
\end{table}

Within the HamLib benchmark, we compare two methods for computing Hamiltonian expectation values, which interact differently with the different GPU parallelism paradigms described above:

\textbf{SpinOperator method:} This method invokes CUDA-Q's native \texttt{observe()} function with a \texttt{SpinOperator} representing the full Hamiltonian. Internally, CUDA-Q computes the expectation value using a single circuit execution that leverages the statevector representation directly, without explicit sampling. This is equivalent to exact classical computation of the expectation value and is most naturally parallelized using \texttt{mgpu} statevector distribution, which reduces execution time by applying the statevector computation across multiple GPUs.

\textbf{Simple sampling method:} This method groups Pauli terms using a commutativity-based strategy, constructs a distinct measurement circuit for each commuting group, and executes these circuits using shot-based sampling. This mimics the execution model of real quantum hardware, where expectation values must be estimated statistically. The multiple circuits produced by this method are the natural target for the \texttt{-{-}pm mpi} parallel execution mode.

An important distinction is that the SpinOperator method, as a form of exact classical computation, is not directly achievable on quantum hardware. It serves as a high-fidelity baseline. The simple sampling method, by contrast, is structurally equivalent to what would be executed on a quantum processor, and the parallelism strategy developed here applies directly to that hardware context. The goal of the sampling mode in this work is not to match or exceed the speed of the SpinOperator computation on the GPU, an outcome that depends on shot count and sampling overhead, but rather to provide a realistic simulation of the hardware execution model and to characterize the speedup achievable through circuit-level parallelism.

The full set of execution configurations evaluated is summarized in Table~\ref{tab:configs}. Four of the configurations represent a $2 \times 2$ design: two observable computation methods (SpinOperator and simple sampling) crossed with two parallelism strategies (single-GPU and multi-GPU). Additionally, we demonstrate the potential of the hybrid parallelism simulations with higher qubit counts. In practice, each configuration is selected through a combination of the \texttt{-g} (grouping method), \texttt{-{-}pm} (parallel mode), and in the case of hybrid parallelism, \texttt{-{-}gpus\_per\_circuit} flags.

\subsection{Four-Configuration Summary at 28 Qubits}

Table~\ref{tab:summary_28q} compares all four execution configurations (two computation methods crossed with two parallelism strategies) at 28 qubits for TFIM and Bose-Hubbard.

\begin{table}[ht]
\centering
\caption{Execution time summary at 28 qubits (16 GPUs). Four configurations: SpinOperator (exact) and simple sampling, each with single-GPU and multi-GPU execution.}
\label{tab:summary_28q}
\begin{tabular}{@{}llrr@{}}
\toprule
\textbf{Hamiltonian} & \textbf{Configuration} & \textbf{Time (s)} & \textbf{Speedup} \\
\midrule
\multirow{4}{*}{TFIM (2 groups)}
  & SpinOperator, 1 GPU    & 0.842 & --   \\
  & SpinOperator, mgpu 16  & 0.607 & 1.4$\times$ \\
  & Simple, 1 GPU          & 1.699 & --   \\
  & Simple, -pm 16         & 0.898 & 1.9$\times$ \\
\midrule
\multirow{4}{*}{Bose-Hubbard (9 groups)}
  & SpinOperator, 1 GPU    & 6.301  & --   \\
  & SpinOperator, mgpu 16  & 3.495  & 1.8$\times$ \\
  & Simple, 1 GPU          & 53.541 & --   \\
  & Simple, -pm 16         & 6.503  & \textbf{8.2}$\times$ \\
\bottomrule
\end{tabular}
\end{table}

For TFIM, the theoretical maximum speedup from circuit-level parallelism is $2\times$ (2 circuits), and we observe $1.9\times$.
For Bose-Hubbard, the $8.2\times$ speedup represents 91\% of the achievable maximum of $\min(16, 9) = 9\times$.
The SpinOperator method, which computes expectation values exactly via statevector manipulation, is inherently faster than shot-based sampling but cannot be replicated on quantum hardware.
It serves as a baseline for comparing the two multi-GPU modes.

\subsection{TFIM: Execution Time vs.\ Qubit Width}

\begin{table}[ht]
\centering
\caption{TFIM Hamiltonian: execution time vs.\ qubit width (16 GPUs).}
\label{tab:tfim}
\resizebox{\columnwidth}{!}{%
\begin{tabular}{@{}rrrrr@{}}
\toprule
& \multicolumn{1}{c}{\textbf{SpinOperator}}
& \multicolumn{3}{c}{\textbf{Simple (2 circuits)}} \\
\cmidrule(lr){2-2}\cmidrule(lr){3-5}
\textbf{Qubits} & \textbf{1 GPU}
& \textbf{1 GPU} & \textbf{-pm 16} & \textbf{Speedup} \\
\midrule
20 & 0.431 & 0.422 & 0.228 & 1.9$\times$ \\
22 & 0.032 & 0.184 & 0.117 & 1.6$\times$ \\
24 & 0.051 & 0.267 & 0.172 & 1.6$\times$ \\
26 & 0.191 & 0.555 & 0.310 & 1.8$\times$ \\
28 & 0.842 & 1.699 & 0.898 & 1.9$\times$ \\
\bottomrule
\end{tabular}
}
\end{table}

The simple sampling speedups are consistently near $2\times$ across all qubit widths, as expected for a 2-circuit workload. Multi-GPU SpinOperator is inefficient at these low qubit counts and by default will not be parallelized across multiple-GPUs below 26 qubits. Therefore, mgpu measurements were not performed with SpinOperator.

\subsection{Bose-Hubbard: Execution Time vs.\ Qubit Width}

\begin{table}[ht]
\centering
\caption{Bose-Hubbard Hamiltonian: execution time vs.\ qubit width (16 GPUs).}
\label{tab:bosehubbard}
\resizebox{\columnwidth}{!}{%
\begin{tabular}{@{}rrrrrrr@{}}
\toprule
 & \multicolumn{3}{c}{\textbf{SpinOperator}} & \multicolumn{3}{c}{\textbf{Simple (9--11 circuits)}} \\
\cmidrule(lr){2-4}\cmidrule(lr){5-7}
\textbf{Qubits} & \textbf{1 GPU} & \textbf{mgpu 16} & \textbf{Speedup}
               & \textbf{1 GPU} & \textbf{-pm 16} & \textbf{Speedup} \\
\midrule
20 & 0.430 & 0.148 & 2.9$\times$ & 0.996  & 0.428 & 2.3$\times$ \\
24 & 0.405 & 0.244 & 1.7$\times$ & 3.636  & 0.805 & 4.5$\times$ \\
28 & 6.301 & 3.495 & 1.8$\times$ & 53.541 & 6.503 & \textbf{8.2}$\times$ \\
\bottomrule
\end{tabular}
}
\end{table}

Parallel circuit execution speedup increases with qubit width (from $2.3\times$ at 20 qubits to $8.2\times$ at 28 qubits) because per-circuit execution time grows exponentially while MPI coordination overhead remains constant.

\subsection{H2: Execution Time vs.\ Qubit Width}

\begin{table}[ht]
\centering
\caption{H2 Hamiltonian: execution time vs.\ qubit width (16 GPUs).}
\label{tab:h2}
\resizebox{\columnwidth}{!}{%
\begin{tabular}{@{}rrrrrrrr@{}}
\toprule
 & & \multicolumn{3}{c}{\textbf{SpinOperator}} & \multicolumn{3}{c}{\textbf{Simple}} \\
\cmidrule(lr){3-5}\cmidrule(lr){6-8}
\textbf{Qubits} & \textbf{Groups} & \textbf{1 GPU} & \textbf{mgpu 16} & \textbf{Sp.}
               & \textbf{1 GPU} & \textbf{-pm 16} & \textbf{Sp.} \\
\midrule
8  & 54   & 0.342  & 0.141  & 2.4$\times$ & 0.917   & 0.230  & 4.0$\times$ \\
12 & 59   & 0.045  & 0.040  & 1.1$\times$ & 1.127   & 0.188  & 6.0$\times$ \\
20 & 1,251 & 0.601  & 0.468  & 1.3$\times$ & 283.814 & 20.506 & \textbf{13.8}$\times$ \\
\bottomrule
\end{tabular}
}
\end{table}

\begin{figure*}[t!]
  \centering
  \includegraphics[width=1.95\columnwidth]{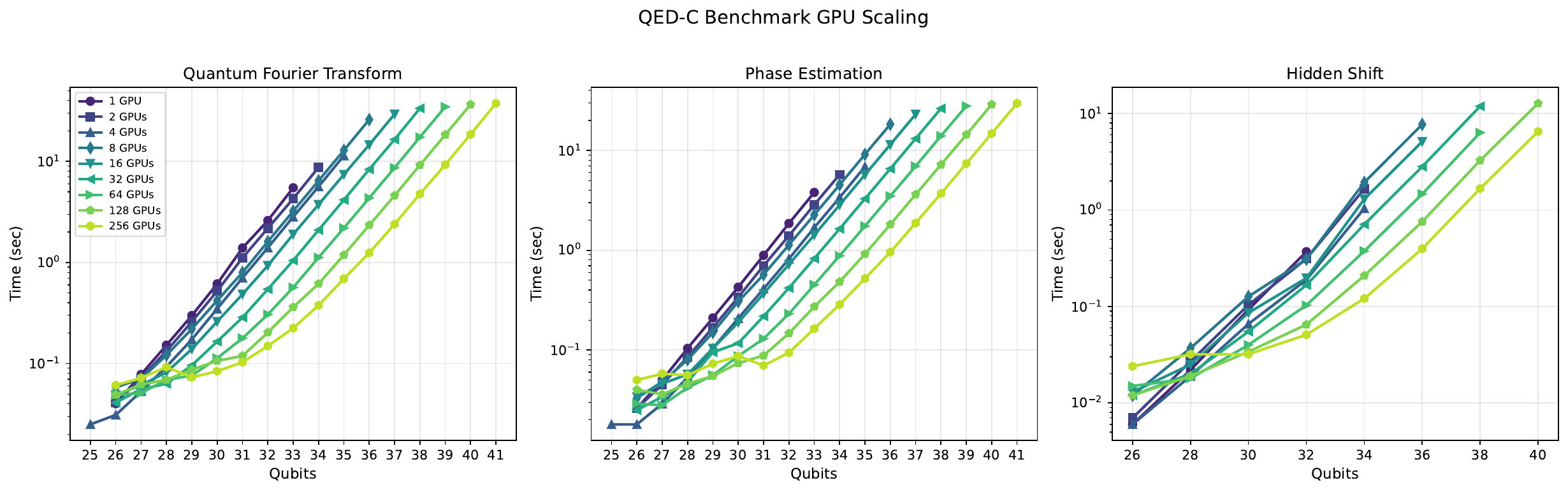}
  \caption{\textbf{GPU scaling behavior for QED-C application-oriented benchmarks.}
Using CUDA-Q \texttt{mgpu} mode on NERSC Perlmutter. Each subplot shows execution time (log scale) versus qubit width for GPU counts ranging from 1 to 256. The three benchmarks---Quantum Fourier Transform (QFT), Quantum Phase Estimation (QPE), and Hidden Shift---each require a single circuit execution whose state vector is distributed across GPU memories. Execution time grows exponentially with qubit count, reflecting $O(2^n)$ state vector scaling. Increasing GPU count shifts the curves downward for large qubit widths and extends the accessible qubit range by $\log_2 G$ for a $G$-GPU configuration.}
  \label{fig:benchmark_scaling}
\end{figure*}

At 20 qubits, 2,951 Pauli terms group into 1,251 independent measurement circuits.
Sequential execution requires 283.8 seconds; with 16 GPUs executing roughly 78 circuits each, wall-clock time reduces to 20.5 seconds ($13.8\times$ speedup, 86\% parallel efficiency).
The SpinOperator method computes the same expectation value in 0.601 seconds on 1 GPU via exact statevector manipulation, illustrating the fundamental speed difference between exact computation and shot-based sampling.
The $13.8\times$ speedup from parallelism substantially narrows this gap.

\subsection{Four-Configuration Comparison}

Figure~\ref{fig:parallel_exec} shows execution time profiles for all four configurations across the three Hamiltonians.

\section{Statevector Distribution Scaling}
\label{app:mgpu_scaling}

This appendix presents results for the \texttt{mgpu} statevector distribution mode, in which all GPUs cooperatively simulate a single circuit by partitioning the statevector across GPU memories.

\subsection{Standard QED-C Benchmark Scaling}

We characterized \texttt{mgpu} scaling across three standard QED-C benchmarks (QFT, QPE, Hidden Shift) on Perlmutter with GPU counts from 1 to 256.
These results establish the baseline scaling behavior of single-circuit distributed simulation.

Figure~\ref{fig:benchmark_scaling} shows execution time as a function of qubit width for each benchmark.
Execution time grows exponentially with qubit count for any fixed GPU configuration, as expected from $O(2^n)$ statevector scaling.
Increasing GPU count shifts the curves downward for larger circuits and extends the accessible qubit range by $\log_2 G$ qubits.
The benefit of additional GPUs is realized primarily at larger qubit widths; for small circuits, MPI communication overhead can exceed the computational savings.

The three benchmarks exhibit broadly similar scaling, consistent with their shared dependence on statevector simulation.
Differences in absolute execution time reflect circuit depth: QFT and QPE ($O(n^2)$ gates) are deeper than Hidden Shift (dominated by Hadamard layers).

\end{document}